\documentclass[usenatbib,useAMS]{mn2e}
\usepackage{natbib}
\usepackage{amsmath}
\usepackage{graphicx}
\usepackage{color}
\usepackage{epsfig}
\allowdisplaybreaks[1]

\usepackage[normalem]{ulem}

\newcommand{\msun}{{\rm M}_\odot}

\newcommand{\zsun}{Z_\odot}
\newcommand{\cc}{{\rm cm}^{-3}}

\newcommand{\msunyr}{{\rm M}_\odot~{\rm yr}^{-1}}

\newcommand{\Hmol}{\rm H_2}
\newcommand{\K}{\rm K}
\newcommand{\beq}{\begin{equation}}
\newcommand{\eeq}{\end{equation}}

\defcitealias{Dijkstra+14}{DFM14}

\title[Suppression of DCBH formation by soft X-rays]
{The suppression of direct collapse black hole formation by soft X-ray irradiation}

\author[K. Inayoshi and T. L. Tanaka]{Kohei Inayoshi$^{1}$\thanks{E-mail: inayoshi@astro.columbia.edu}, Takamitsu L. Tanaka$^{2,3,4}$\thanks{E-mail: takamitsu.tanaka@stonybrook.edu}\\
$^{1}$Department of Astronomy, Columbia University, 550 W. 120th Street, New York, NY 10027, USA\\
$^{2}$Department of Physics and Astronomy, Stony Brook University, Stony Brook, NY 11794, USA\\
$^{3}$Department of Physics, New York University, 4 Washington Place, New York, NY 10003, USA\\
$^{4}$Max Planck Institute for Astrophysics, Karl-Schwarzschild-Str. 1, D-85741 Garching, Germany
}

\begin{document}
\maketitle
\label{firstpage}

\begin{abstract}The origin of supermassive black holes (SMBHs) in galactic nuclei
is one of the major unsolved problems in astrophysics.
One hypothesis is that they grew from $\ga 10^{5} \, \msun$ black holes
that formed in the `direct collapse' of massive
gas clouds that have low concentrations of both metals and molecular hydrogen ($\Hmol$).
Such clouds could form in the early ($z\ga 10$) Universe if pre-galactic gas is irradiated by
$\Hmol$-photodissociating, far-ultraviolet (FUV) light from a nearby star-forming galaxy.
In this work, we re-examine the critical FUV flux $J_{\rm crit}$ that is required
to keep $\Hmol$ photodissociated and lead to direct collapse.
We submit that the same galaxies that putatively supply the extraordinary 
FUV fluxes required for direct collapse should also
produce copious amounts of soft X-rays,
which work to offset $\Hmol$ photodissociation by increasing the ionization fraction
and promoting $\Hmol$ formation.
Accounting for this effect increases the value of $J_{\rm crit}$ by a factor of at least 
$3-10$, depending on the brightness temperature of FUV radiation.
This enhancement of $J_{\rm crit}$ suppresses the abundance of potential direct collapse 
sites at $z>10$ by several orders of magnitude.
Recent studies---without accounting for the soft X-rays from
the FUV source galaxies---had already arrived at large values of $J_{\rm crit}$
that implied that direct collapse may occur too rarely
to account for the observed abundance of high-redshift quasars.
Our results suggest that $J_{\rm crit}$ should be even higher
than previously estimated,
and pose an additional challenge for the direct collapse scenario via strong FUV radiation
to explain the high-redshift quasar population.
\end{abstract}


\section{Introduction}

Most nearby massive galaxies harbour
a supermassive black hole (SMBH) in their nuclei.
Empirical correlations between the masses of SMBHs and
properties of their host galaxies suggest that SMBHs
may play a key role in galaxy evolution,
possibly during stages shining as luminous quasars
\citep[e.g.][]{1998AJ....115.2285M, 2000ApJ...539L...9F,
2003ApJ...589L..21M,2007ApJ...669...67H,2013ARA&A..51..511K}.
Despite their apparent ubiquity and importance,
when and how these cosmic behemoths formed
remain poorly understood.
Observations of luminous quasars at $z\ga 6$ reveal that SMBHs with masses of 
$\ga 10^9~\msun$ were already in place $900$ Myr after the Big Bang,
and place strong constraints on possible formation scenarios
\citep{Fan2001,2006NewAR..50..665F,2010AJ....139..906W,2011Natur.474..616M,
2013ApJ...779...24V,2014AJ....148...14B}.

One possibility is that the earliest SMBHs grew from `seed' $\sim 100~\msun$ BHs
left behind by the first generation of stars (Population III or `Pop III' stars)
from $z\ga 30$ via rapid gas accretion, aided by hierarchical BH mergers
\citep{HaimanLoeb01, MadauRees01, VHM03, Li+07, TPH12, TLH13}\footnote{
BH mergers play a secondary or minor role in BH growth. The gravitational recoil effect,
while unlikely to prevent SMBH formation, suppresses the efficacy of mergers in assembling
more massive BHs \citep{VR06, TH09}.}.
To form the $z>6$ quasar SMBHs, these seeds must have accreted gas at
a mean rate comparable to the Eddington limit; a key uncertainty is whether
Pop III remnant BHs could have maintained such rates despite negative radiative feedback
in the shallow gravitational potentials of their host protogalaxies
\citep{2009ApJ...701L.133A,2009ApJ...696L.146M}.

Alternatively, SMBHs could have originated as $\ga 10^{5-6}~\msun$ BHs that formed
via the `direct collapse' of gas clouds with low abundances of metals and molecular hydrogen ($\Hmol$)
\citep[e.g.][]{1994ApJ...432...52L, 2002ApJ...569..558O,
2003ApJ...596...34B, 2004MNRAS.354..292K, 2006MNRAS.370..289B,2006MNRAS.371.1813L}.
Theoretically, direct collapse can occur in primordial gas clouds in massive dark-matter halos 
with virial temperatures of $\ga 10^4~\K$,
if $\Hmol$-line cooling is suppressed.
The most widely studied $\Hmol$-suppressing mechanism in this context is
photodissociation by strong far-ultraviolet (FUV) radiation in the Lyman-Werner (LW) band ($11.2-13.6$ eV)
\footnote{For alternative mechanisms, see \cite{IO12}, \cite{TL14}.}.
In such primordial gas without H$_2$ molecules,
the gas loses thermal energy primarily via atomic hydrogen transitions
(Ly$\alpha$, two-photon, and H$^-$ free-bound emissions)
and collapses while maintaining a temperature of $\sim 8000$ K \citep[e.g.,][]{O01}.
Recent numerical simulations have suggested that the gas can collapse monolithically avoiding
the major-episode of fragmentation during the isothermal phase 
\citep{2010MNRAS.402.1249S, 2009MNRAS.396..343R, 2009MNRAS.393..858R,
2013MNRAS.433.1607L, 2014MNRAS.445L.109I,2015MNRAS.446.2380B}.

After the collapse phase, a protostar with a mass of $\sim 1~\msun$ forms at the centre of the cloud
and grows via rapid gas accretion at the rate of $\ga 1~\msunyr$ \citep{2014MNRAS.445L.109I}.
The protostar growing at such a high accretion rate evolves to a supermassive star within its lifetime 
$\sim 1$ Myr
overcoming the radiative feedback and pulsation-driven mass loss
\citep{2012ApJ...756...93H, 2013MNRAS.431.3036I, 2013ApJ...778..178H, 2013A&A...558A..59S}
and finally forms a massive seed BH by 
gravitational collapse due to general relativistic instability 
\citep{1964ApJ...140..417C,1971reas.book.....Z,2002ApJ...572L..39S}.
Compared to Pop III seed BHs, the products of direct collapse 
(`direct collapse black holes,' henceforth DCBH)
require a somewhat lower (by $\sim 10$ to $20$ per cent)
mean accretion rate to grow to $\ga 10^9~\msun$ by $z\sim 6-7$
(although the rate is still comparable to the Eddington limit; see e.g. 
\citealt{Tanaka14} and references therein).

The most crucial question in the above scenario is
how large the LW intensity must be to 
keep $\Hmol$ dissociated.
This critical value, commonly called $J_{\rm crit}$,
has been discussed by many authors
\citep{O01,2003ApJ...596...34B,2010MNRAS.402.1249S,IO11,2011MNRAS.418..838W,
2014MNRAS.443.1979L}.
If the irradiating source has a thermal spectrum with a brightness temperature $T_\ast$,
$J_{\rm crit}\simeq \mathcal{O}(10)$
(in units of $10^{-21}$ erg s$^{-1}$ cm$^{-2}$ sr$^{-1}$ Hz$^{-1}$) 
for $T_\ast =10^4$ K and $J_{\rm crit}\simeq \mathcal{O}(10^3)$ for $T_\ast =10^5$ K.
For example, \cite{2014MNRAS.445..544S} recently obtained  $J_{\rm crit}\simeq 1400$
and found that this value does not change significantly between realistic UV spectra
of star-forming, low-metallicity galaxies.
Several studies have estimated the probability of forming DCBHs via FUV fluxes $J_{\rm LW}>J_{\rm crit}$,
using Monte Carlo calculations \citep[][hereafter, DFM14]{Dijkstra+08, Dijkstra+14}
and semi-analytic methods coupled with $N$-body simulations 
\citep{Agarwal+12}.
If $J_{\rm crit}\ga 10^{3}$, then the expected number density of DCBHs
is comparable to or lower than that of SMBHs with $\ga 10^9~\msun$ 
at $z\ga 6$ ($\sim 1$ comoving ${\rm Gpc}^{-3}$).
Recently, \cite{Latif+14} found that $J_{\rm crit}$ may be as large as $>10^4$,
further challenging the viability of the DCBH model to explain the
observed SMBH population.

In this paper, we discuss the role of X-rays from the same star-forming galaxies
that are the putative sources of $\Hmol$-dissociating FUV radiation.
X-rays can increase the hydrogen ionization fraction, promoting $\Hmol$ formation
through the electron-catalyzed reactions
\begin{align}
{\rm H} + {\rm e}^- &\rightarrow {\rm H}^- + \gamma ,
\label{eq:H2react1}\\
{\rm H} ^- + {\rm H} &\rightarrow {\rm H}_2 + {\rm e}^-.
\label{eq:H2react2}
\end{align}
By working to \textit{increase} the $\Hmol$ fraction,
X-rays work against FUV photons and thus
increase the effective value of $J_{\rm crit}$---that is, we should generally expect
\begin{equation}
J_{\rm crit}^{\rm (UV+X)} > J_{\rm crit} ^{\rm (UV~only)}
\end{equation}
if nearby galaxies irradiate putative DCBH formation sites with FUV \textit{and} X-ray photons \citep{IO11}.
The above suggests that failing to account for the ($\Hmol$-promoting) X-ray intensity
that accompanies the ($\Hmol$-dissociating) FUV intensity will generally result in 
an underestimate of $J_{\rm crit}$.

That additional ionization sources in general (e.g. cosmic rays) can 
increase $J_{\rm crit}$ was made by \cite{IO11}.
Recently, \cite{Latif+14} investigated the effect of
$E\ge 2$ keV (hard) X-rays, which form a cosmic background
because their mean free paths are too long to be absorbed locally.
They found that the role of the X-rays become non-negligible at intensities $J_{\rm X,21}>0.01$,
a value much higher than the expected background at $z>10$.
In this work, we invoke observational results from lower redshifts to
argue that star-forming galaxies that supposedly irradiate
putative DCBH formation sites with $J_{\rm LW,21}\ga 2\times 10^3$
should also supply a soft ($\la 1$ keV) X-ray intensity $J_{\rm X,21}\ga 0.01$,
We show that such intensities of soft local X-rays can raise $J_{\rm crit}$,
just as the previous works \citep{IO11,Latif+14} found for hard background X-rays.

For the most conservative assumption that
high-redshift star-forming galaxies produce the same ratio of X-ray to FUV photons
as local star-forming galaxies, accounting for X-ray ionizations increases $J_{\rm crit}$
by a factor of at least $\sim 3-10$ compared to previous work.
This increase is highly sensitive to the actual X-ray to FUV flux ratio,
and can be larger than an order of magnitude if early galaxies produce more X-rays
relative to FUV photons.
Following the semianalytic methods of \citetalias{Dijkstra+14},
we show that even a modest increase in $J_{\rm crit}$ 
reduces the abundance of DCBH formation sites by several orders of magnitude,
further challenging the viability of the DCBH scenario.

We stress that this effect holds regardless of theoretical uncertainties
in the calculation of the X-ray-uncorrected critical flux $J_{\rm crit}^{\rm (UV~only)}$.
Whether  $J_{\rm crit}^{\rm (UV~only)}\sim 10^3$ as suggested by one-zone
calculations, or $\sim 10^4$ as suggested by three-dimensional simulations,
soft X-rays ($\sim 1$ keV) from the FUV source galaxies will act to increase $J_{\rm crit}$.
In other words, the primary goal of this work is not to claim that $J_{\rm crit}$ 
may be too high to explain the observed SMBH abundance
(as this has already been suggested by \citealt{Latif+14} and others),
but rather to demonstrate that it should be \textit{higher} than found
by previous studies that did not account for the soft X-ray output
of the FUV source galaxies.

The rest of this paper is organised as follows.
We describe in \S\ref{sec:method}
our calculations of the critical LW intensity,
in particular our treatment of X-ray ionization.
In \S\ref{sec:Jcrit}, we quantify the relation between
LW and X-ray radiation from star-forming galaxies in the early Universe,
and arrive at a relationship between the UV-only and X-ray-corrected
values of $J_{\rm crit}$ ($J_{\rm crit}^{\rm (UV~only)}$ and $J_{\rm crit}^{\rm (UV+X)}$, respectively).
In \S\ref{sec:prob}, we apply these results to arrive at 
the X-ray-corrected probability that an atomic-cooling halo can form a DCBH.
We estimate the number density of DCBHs as a function of $J_{\rm crit}$ and redshift.
Finally, we present our conclusions in \S\ref{sec:conc} and discuss the 
potential role of 21cm signatures and other observations in placing
empirical constraints on FUV-aided DCBH formation.

\section{Evaluation of $J_{\rm crit}$}
\label{sec:method}

\subsection{Thermal and chemical evolution}
We consider the thermal evolution of primordial gas in a massive halo with 
a virial temperature of $\ga 10^4$ K that is exposed to FUV radiation and X-rays from external sources.
During the collapse of the self-gravitating cloud, its density profile approaches 
a self-similar form \citep{1969MNRAS.144..425P,1969MNRAS.145..271L},
consisting of a central core and an envelope with $\rho \propto r^{-2}$.
We here adopt a one-zone model which approximates
all the physical quantities to be uniform inside
the central core, and solve for their temporal evolution \citep[e.g.,][]{O01}.

The density of the central core increases on the free-fall timescale
$t_{\rm ff}=\sqrt{3\pi/32G\rho}$ as
\begin{equation}
\frac{{\rm d}\rho}{{\rm d}t}=\frac{\rho}{t_{\rm ff}}.
\end{equation}
The energy equation of the gas is given by
\begin{equation}
\frac{{\rm d}e}{{\rm d}t}=-p\frac{\rm d}{{\rm d}t}\left(\frac{1}{\rho}\right)-\frac{\Lambda - \Gamma_{\rm X}}{\rho},
\end{equation}
where $e$ is the specific internal energy, $p$ the gas pressure, $\Lambda$ the cooling rate, and
$\Gamma_{\rm X}$ the heating rate due to the external X-rays.
We consider the radiative cooling by
atomic and molecular hydrogen species, as well as
the cooling/heating associated with chemical reactions.
As the collapse proceeds and the gas grows denser,
the intensity of external radiation that reaches the central core is reduced.
We estimate the optical depth by assuming the size of the central core
to be the half of the Jeans length $\lambda_{\rm J}$.
At the collapsing central core, the column density of the $i$-th species is given by 
\begin{equation}
N_{\rm i}=n(i)\frac{\lambda_{\rm J}}{2},
\end{equation}
where $n(i)$ is the number density of the species.

We solve the primordial chemical reactions among the following 9 species:
H, H$_2$, e$^-$, H$^+$, H$_2^+$, H$^-$, He, He$^+$, and He$^{++}$.
The chemical reactions we consider are the same as in \cite{O01}
but we have updated some reaction rate coefficients \citep{2014MNRAS.445L.109I}.
We include the photoionization of H and He by X-rays.

The one-zone calculations start at $n=0.1~\cc$ and $T=160$ K, which corresponds to the gas in
a halo virializing at $z_{\rm vir}\simeq 10$ \citep{2008ApJ...686..801O}.
We set the initial abundances of electrons, H$_2$, and He to $x_{\rm e}=10^{-4}$, 
$x_{\rm H_2}=10^{-6}$, and $x_{\rm He}=0.08$, respectively.
These initial conditions are the same as in \cite{IO11}.

\subsection{External FUV and X-ray radiation}
We now discuss our treatment of FUV and X-ray radiation,
and in particular how X-rays affect the effective value of $J_{\rm crit}$.
Below, we use the symbol $J_{\rm crit,0}\equiv J_{\rm crit}^{\rm (UV~only)}$
to denote the value of $J_{\rm crit}$ calculated without considering the effects of X-ray ionizations.

\subsubsection{FUV radiation}
We assume the FUV radiation to have a diluted thermal spectrum,
$J_{\rm LW}(\nu)\propto B_\nu(T_\ast)$, and consider brightness temperatures of 
$T_\ast =2\times 10^4$, $3\times 10^4$ and $10^5$ K.
These values of $T_\ast$ correspond to realistic spectra of Pop II/III star-forming galaxies,
the FUV sources near DCBH forming halos (see \S\ref{seq:FUV_rad}).
We normalise the intensity of the FUV radiation at the Lyman limit ($\nu_{\rm L}=13.6$ eV),
and write this in conventional units:
$J_{\rm LW,21}=J_{\rm LW}(\nu_{\rm L})/(10^{-21}~{\rm erg}~{\rm s}^{-1}~{\rm cm}^{-2}
~{\rm sr}^{-1}~{\rm Hz}^{-1})$. 
For the thermal spectra we adopt, the rate coefficients for photodissociation
of  H$_2$ and H$^-$ are
given by $k_{\rm H_2}^{\rm pd}\equiv \kappa_{\rm H_2}J_{\rm LW,21}$
and $k_{\rm H^-}^{\rm pd}\equiv \kappa_{\rm H^-}J_{\rm LW,21}$, respectively.
The values of $\kappa_{\rm H_2(H^-)}$ for each $T_\ast$ are listed in Table ~\ref{tab_phch}.
We have also included the case $T_\ast=10^4$ K for reference.
We also consider the H$_2$ self-shielding effect against external
FUV radiation \citep{2011MNRAS.418..838W}.

\begin{table}
\begin{center}
\caption{Photodissociation rates of H$_2$ and H$^-$ (in cgs units) for thermal spectra with 
brightness temperature $T_\ast$.}
  \begin{tabular}{c|c|c|c|c} \hline
$T_\ast$ (K) & $2\times 10^4$ & $3\times 10^4$ & $10^5$ & ($10^4$)\\ \hline
\vspace{2mm}
$\kappa_{\rm H_2}\times 10^{12}$  & $2.1$ & $1.7$ & $1.3$ & $4.3$ \\
\vspace{2mm}
$\kappa_{\rm H^{-}}\times 10^{10}$  & $4.4$ & $0.77$ & $0.13$ & $2000$\\
$k_{\rm H^-}^{\rm pd}/k_{\rm H_2}^{\rm pd}$ & $2.1\times 10^2$& $46$ & 
$10$ & $4.6\times 10^4$\\\hline
   \end{tabular}
   \label{tab_phch}
  \end{center}
\end{table}

\subsubsection{X-rays}
We assume that the X-ray mean intensity can be represented by a power-law spectrum,
\begin{equation}
J_{\rm X}(\nu)=J_{\rm X,21}\times 10^{-21}\left(\frac{\nu}{\nu_0}\right)^{-\alpha}
~{\rm erg}~{\rm s}^{-1}~{\rm cm}^{-2}~{\rm sr}^{-1}~{\rm Hz}^{-1},
\label{eq:spect}
\end{equation}
where $h\nu_0=1$ keV and $\alpha =1.8$ \citep[e.g.][]{2004ApJS..154..519S}.
The ionization rates of H and He by direct X-ray photons are given by  
\begin{align}
\zeta_{\rm X,p}^i &=\int^{\nu_{\rm max}}_{\nu_{\rm min}}\frac{4\pi J_{\rm X}(\nu)}{h\nu}
e^{-\tau_{\nu}}\sigma_i(\nu) d\nu~~~(i={\rm H},~{\rm He}),\\
\tau_\nu &=N_{\rm H}\sigma_{\rm H}(\nu)+N_{\rm He}\sigma_{\rm He}(\nu),
\end{align}
where $\sigma_{\rm H}(\nu)$ and $\sigma_{\rm He}(\nu)$ are the cross sections 
of H and He to the ionizing photons 
\citep[][respectively]{1996ApJ...465..487V, 1998ApJ...496.1044Y}
\footnote{Previously, \cite{IO11} adopted the cross sections by \cite{1979rpa..book.....R} (for H)
and \cite{1989agna.book.....O} (for He).
These works overestimate the cross sections at X-ray energies.
In this work we adopt the more recent cross sections referenced above.}
and $N_{\rm H}$ and $N_{\rm He}$ are the column densities of those species.
Since the emitted electrons have large kinetic energy, they can ionise the surrounding gas
(secondary ionization).
We also estimate the secondary ionization and X-ray heating rates
using the formulae in \cite{1985ApJ...298..268S}, which are valid for
X-ray photons with energies $\gg 0.1$ keV.

In this calculation, we set the maximum energy of the X-rays to $h\nu_{\rm max}=10$ keV.
The following results do not depend on
the choice of $\nu_{\rm max}$ as long as $h\nu_{\rm max}\geq 10$ keV.
The X-ray minimum energy is the more important quantity.
The comoving mean free path of a X-ray photon with $h\nu$ can be written as
\begin{equation}
\lambda_{\rm X }\simeq 9.1 {\bar x_{\rm H}}^{-1}\left(\frac{1+z}{11}\right)^{-2}
\left(\frac{h\nu}{0.3~{\rm keV}}\right)^3~{\rm cMpc},
\end{equation}
where ${\bar x_{\rm H}}$ is the mean neutral fraction \citep{2006PhR...433..181F}.
From the condition that $\lambda_{\rm X}$ is longer than the Hubble horizon, 
hard X-ray photons with $\ga 2$ keV can build up a cosmic X-ray background 
before $z\sim10$ \citep[e.g.][]{RicottiOstriker04,TPH12}.
On the other hand, the FUV sources required to form DCBHs are star-forming galaxies close to the 
DCBH forming halo.
The physical separation is typically $\sim 10$ kpc \citep{Dijkstra+08}, 
which is much shorter than $\lambda_{\rm X}/(1+z)\sim 800$ kpc.
We therefore argue that soft X-rays with energies $h\nu<2$ keV,
if produced by the FUV sources, must irradiate potential DCBH formation sites.
Given the uncertainty in the X-ray emission properties
of the earliest star-forming galaxies, we consider minimum X-ray energies of 
$h\nu_{\rm min}=0.5$ and $1$ keV.
For comparison to previous works,
we also consider the case $h\nu_{\rm min}=2$ keV.

\subsection{X-ray enhancement of the critical LW flux}
\label{sec:2.3}
\begin{figure}
\begin{center}
\includegraphics[height=60mm,width=80mm]{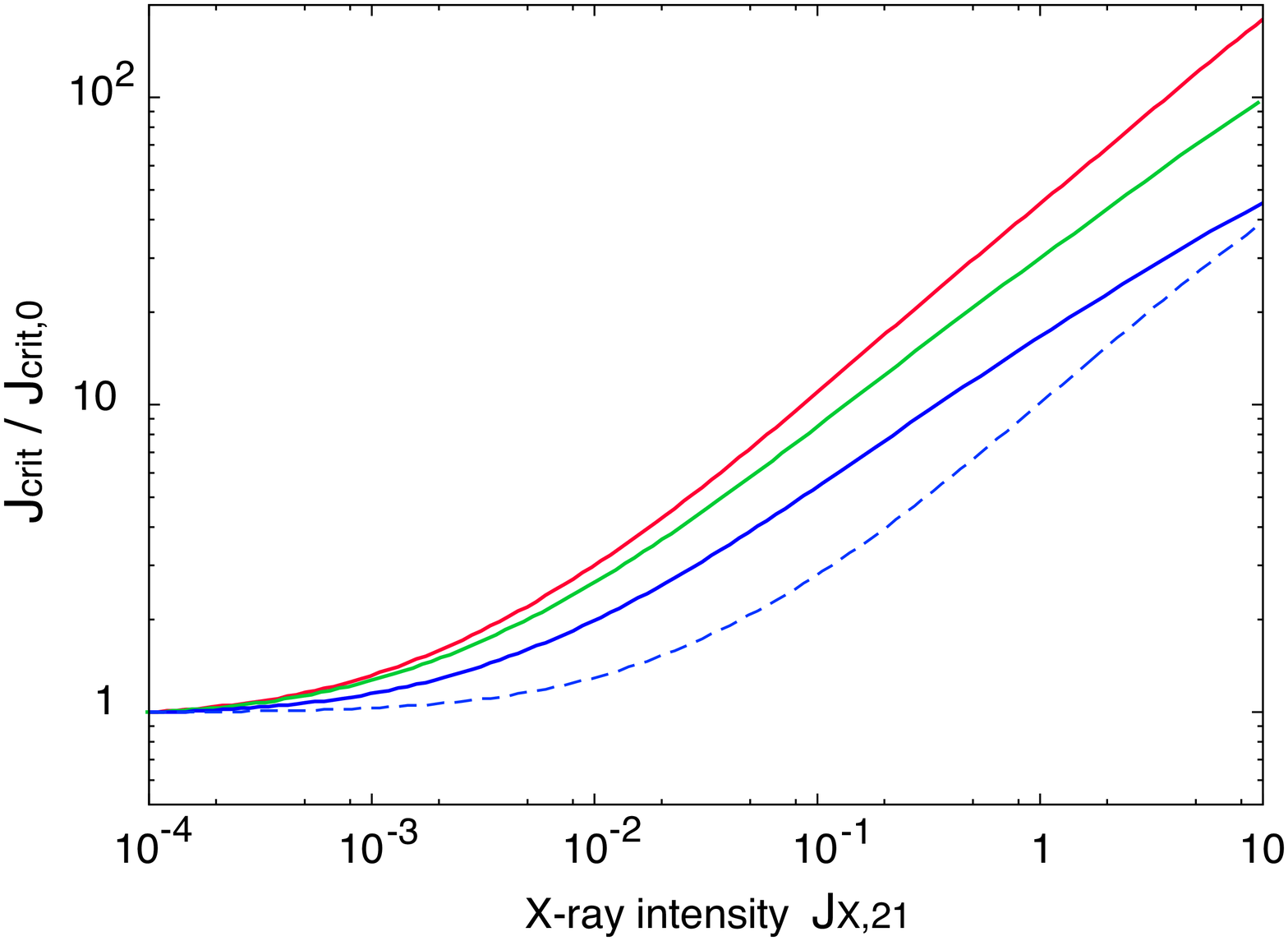}
\end{center}
\caption{The ratio $J_{\rm crit}/J_{\rm crit,0}$, the relative increase in the requisite 
FUV intensity to photodissociate H$_2$ molecules inside an atomic-cooling halo, 
when X-rays are added to the radiating spectrum. 
The solid lines show the cases for $h\nu_{\rm min} = 1$ keV;
from bottom to top:
$T_\ast =2\times 10^4$ (blue curve), $3\times 10^4$ (green) and $10^5$ K (red).
The results for $h\nu_{\rm min} = 0.5$ keV are nearly identical to those
for $h\nu_{\rm min} = 1$ keV and we do not show them here.
The dashed line shows the case of $h\nu_{\rm min} = 2$ keV and 
$T_\ast =2\times 10^4$ K, for comparison with the previous study.
A colour version of this figure is available in the online version.
}
\label{fig:1}
\end{figure}

In Fig.~\ref{fig:1}, we present the ratio of $J_{\rm crit}/J_{\rm crit,0}$ (FUV+X to FUV only)
as a function  of the X-ray intensity $J_{\rm X,21}$.
The solid curves show cases with $h\nu_{\rm min}\leq 1$ keV 
and $T_\ast =2\times 10^4$ (blue), $3\times 10^4$ (green), and $10^5$ K (red) from bottom to top.
The dashed blue curve shows the case  $h\nu_{\rm min}=2$ keV
and $T_\ast =2\times 10^4$ K, for comparison with a previous study \citep[][see below]{Latif+14}.

Let us first discuss the case $h\nu_{\rm min}=1$ keV.
For weak X-ray intensities, the value of $J_{\rm crit}/J_{\rm crit,0}$
converges to a constant value of $\approx 1$.
The critical LW intensity increases with the X-ray intensity at $J_{\rm X,21}\ga 10^{-3}$ 
for all cases of $T_\ast$.
This is because the electron fraction rises through X-ray ionization and 
thus the H$_2$ formation rate through the reactions in equation~(\ref{eq:H2react1}) 
and (\ref{eq:H2react2}) increases
(We show the evolution of the electron fraction and discuss 
the effect of X-ray ionization on its behavior in the Appendix.).
For all cases considered here, the functional form of $J_{\rm crit}$
can be fitted as $J_{\rm crit}/J_{\rm crit,0}=(1+J_{\rm X,21}/a)^b$,
where $(J_{\rm crit,0},~a,~b)=(1.7\times 10^3,~3.4\times 10^{-3},~0.50)$, 
$(2.4\times 10^3,~2.2\times 10^{-3},~0.56)$ and 
$(3.1\times 10^3,~2.1\times 10^{-3},~0.62)$ for 
$T_\ast =2\times 10^4$, $3\times 10^4$
and $10^5$ K, respectively
\footnote{
If we normalise the FUV intensity at the average energy of the LW band (12.4 eV) instead of 
$\nu_{\rm L}=13.6~{\rm eV}$, the values of $J_{\rm crit,0}$ become a constant ($\simeq 2.7\times 10^3$) 
for all cases of $2\times 10^4\leq T_\ast \leq 10^5$ K.
This dependence on $T_\ast$ agrees  with previous work \citep{2014MNRAS.445..544S}.}

Next, we address the dependence of the critical FUV intensity on 
the X-ray intensity and the FUV spectrum. 
As many authors have shown \citep[e.g.,][]{O01,IO11}, 
the value of $J_{\rm crit,0}$ for hard spectra ($T_\ast \ga 2\times 10^4$ K)
is well-approximated by a functional form $f(T_\ast) x_{\rm e}$,
because H$^-$ photodissociation is negligible for $J_{\rm crit,0}$ in such harder spectra
\citep[see][for details]{IO11}.
That is, the value of $J_{\rm crit,0}$ scales linearly with the electron fraction and 
the normalization depends on the FUV spectrum. 
The primary role of the X-rays is to raise $J_{\rm crit}$ by increasing the 
equilibrium value of $x_{\rm e}$.
For simplicity, if we assume that the X-ray ionization and recombination are balanced 
in the collapsing cloud, the electron fraction is expressed as $x_{\rm e}\propto J_{\rm X,21}^{1/2}$.
This approximation is valid for $n_{\rm H}\la 10^4~\cc$ and explains why the power-low index
$b\simeq 0.5$ for the fitting form described in the previous paragraph. 
Thus, $J_{\rm crit}/J_{\rm crit,0}$ can be approximated by the ratio of 
the values of $x_{\rm e}$ obtained with and without X-ray
ionizations---that is, it depends most sensitively on the X-ray intensity, 
and depends weakly on the choice of FUV spectrum 
as long as H$^-$ photodissociation is negligible.
We provide extended descriptions of the above relationships in the Appendix,
and direct readers there for additional details.

The values of $J_{\rm crit}/J_{\rm crit,0}$ for $T_\ast =3\times 10^4$ K are indeed
close to those for $T_\ast =10^5$ K.
However, for a high value of $J_{\rm crit}$, 
the {\it indirect} H$_2$ dissociation through the H$^-$ photodissociation begins to work 
against H$_2$ formation by the reaction in equation (\ref{eq:H2react2})---i.e.
$\kappa_{\rm H^-}J_{\rm crit}>k_2 n_{\rm H2}^{\rm crit}
\simeq 1.1\times 10^{-6}$ at $T\simeq 8000$ K,
where $k_2$ is the rate coefficient of the reaction (\ref{eq:H2react2})
and $n_{\rm H2}^{\rm crit}$ is the critical density of H$_2$.
The above condition is written as $J_{\rm crit}/J_{\rm crit,0}>1.5$, $6.0$, and $27$
for $T_\ast = 2\times 10^4$ K, $3\times 10^4$ K and $10^5$ K, respectively.
Therefore, for larger $J_{\rm crit}$, the increase of $J_{\rm crit}/J_{\rm crit,0}$ becomes 
slightly smaller. In Fig.~\ref{fig:1}, this can be seen in the decrease in the slope of this
ratio at large $J_{\rm X,21}$ for the $T_\ast = 2\times 10^4$ K case.
However, we note that this effect is smaller than many other uncertainties in evaluating $J_{\rm crit}$
(see \S\ref{subsec:actual_jcr}).

We study the dependence of the critical LW intensity $J_{\rm crit}$ on the choice of $h\nu_{\rm min}$.
In the case of $h\nu_{\rm min}=2$ keV, the minimum X-ray intensity
for increasing $J_{\rm crit}$ is larger by one order of magnitude
than that for our fiducial model ($h\nu_{\rm min}=$ 1 keV).
This is because the cross sections of H and He against
photons with energies $\ga 2$ keV are small and 
thus the X-ray ionization is also less important.
For this case, the fitted parameters are given by
$(J_{\rm crit,0},~a,~b)=(1.7\times 10^3,~2.2\times 10^{-2},~0.6)$.
The curve for $h\nu_{\rm min}=2$ keV shifts rightwards by approximately
one order of magnitude compared to that for $h\nu_{\rm min}=1$ keV.

Recently, \cite{Latif+14} studied the effect of X-rays on DCBH formation by considering hard X-rays
($h\nu_{\rm min}=2$ keV) that comprise the X-ray background radiation.
In their result, the critical FUV intensity begins to increase at $J_{\rm X,21}\ga 10^{-2}$ 
and boosts by a factor of $\approx 2$ at $J_{\rm X,21}\simeq 10^{-1}$.
This is consistent with our findings, represented by the blue dashed curve
in Fig.~\ref{fig:1}. Note that this result is robust to changes in $T_\ast$, as explained above.
As seen the dashed line (blue) in Fig.~\ref{fig:1}, our result is $J_{\rm crit}/J_{\rm crit,0}\approx 2.5$
for $J_{\rm X,21}\simeq 10^{-1}$, which is consistent with their three-dimensional simulation.
A minimum energy of $h\nu_{\rm min}=2$ keV
is a reasonable assumption for the cosmic X-ray background,
as soft X-rays with $0.5~(1)$ keV would be absorbed at separations of $\sim 40~(300)~{\rm Mpc}$ 
in the intergalactic medium.
However, the {\it nearby} halos that putatively enable DCBH formation
through large FUV fluxes should also irradiate their immediate environments
with soft ($\sim 1~{\rm keV}$) X-rays.
{\it We emphasize the key point that soft X-rays are far more effective at promoting
$\Hmol$ formation} through electron-catalyzed reactions.

On the other hand, the result for $h\nu_{\rm min}=0.5$ keV does not change from 
our fiducial model ($h\nu_{\rm min}=1$ keV).
The value of $J_{\rm crit}$ is determined by the electron fraction at $n_{\rm H}\simeq 10^3~\cc$ 
and $T\simeq 8000$ K \citep{IO11}.
The photons with $h\nu_{\rm min}\la 1$ keV can ionise the gas easily and thus are absorbed 
at $n_{\rm H}<10^3~\cc$.
We conclude that the value of $J_{\rm crit}$
is sensitive to the intensity of X-ray photons at $\simeq 1~{\rm keV}$
but not to that of softer X-rays at energies $\la 1~{\rm keV}$.

Here, we have assumed a simple power-law spectrum with $J_{\rm X}(\nu)\propto \nu ^{-1.8}$.
However, the spectral energy distributions of observed high-mass X-ray binaries (HMXBs) 
are more complex 
\citep[e.g.,][]{1997MNRAS.288..958G,1999MNRAS.309..496G,2004ApJS..154..519S}.
The spectral shapes are characterised
by a power-law with $\alpha \approx 1.6-1.8$ (low-hard state)
and by a bright thermal component with a peak temperature of $\sim 0.5$ keV
having a soft power-law tail with $\alpha \ga 2.0$ (high-soft state).
We note that the resulting value of $J_{\rm crit}/J_{\rm crit,0}$ for $h\nu_{\rm min}\leq1$ keV 
depends very weakly on the choice of the X-ray power-law index in the range of $1.6\la \alpha \la 2.0$
because ionization by soft X-rays ($\sim 1$ keV) increases $J_{\rm crit}$ significantly.
The X-ray spectra of HMXBs
in high-$z$ galaxies could have an excess due to 
the thermal emissions from the power-law component 
at $\sim 1-10$ keV \citep{2013ApJ...776L..31F}.
We note that the value of $J_{\rm crit}$ begins to increase 
for X-ray intensities as small as $J_{\rm X,21}\sim 10^{-3}$
in a case with thermal soft X-ray components.

\section{LW and X-ray sources in the early Universe}
\label{sec:Jcrit}

Having laid out above the general effect of X-ray fluxes
on the quantity $J_{\rm crit}$, we now turn to the discussion
of X-ray and LW sources in the $z\ga10$ Universe.

\subsection{FUV and X-ray intensities}
\subsubsection{X-ray flux}
\label{sec:Xray-flux}

We first estimate the X-ray intensities from the star-forming galaxies
in the $z\sim 20-10$ Universe. 
According to the most recent cosmological simulations,
Pop III stars could be born as massive stars with $\sim 10-100~\msun$ 
\citep[e.g.,][]{2011Sci...334.1250H, 2012MNRAS.422..290S, 2014ApJ...781...60H}. 
Moreover, the efficiency of forming binary systems could be as high as $\sim 50~\%$ \citep[e.g.,][]{2013MNRAS.433.1094S,2014ApJ...792...32S}.
Thus, we can consider HMXBs as X-ray sources in the early Universe
\citep[e.g.][]{2009MNRAS.395.1146P,2011A&A...528A.149M}
\footnote{Another candidate is a supernova remnant where accelerated electrons emit X-ray photons.
However, the X-ray from supernova remnants could be subdominant 
because of their transient nature
\citep{2001ApJ...553..499O, 2006MNRAS.371..867F,2011A&A...528A.149M}.}.

From observations of local starburst galaxies, we can obtain 
a good correlation between the X-ray luminosities and their star formation rate (SFR).
The X-ray emission is dominated by HMXBs, which is considered to
be good tracers of the SFR because of their short lifetime.
The bolometric X-ray luminosity (2--10 keV) is given by
\begin{equation}
L_{\rm 2-10keV}\simeq 6.7\times 10^{39}\left(\frac{{\rm SFR}}{\msunyr}\right)~{\rm erg~s}^{-1},
\label{eq:LX_SFR}
\end{equation}
\citep[e.g.][]{2003MNRAS.339..793G,2003MNRAS.340..210G}.
Many observations in various X-ray bands also have suggested the same $L_{\rm X}$-SFR relation
within a factor of $2-3$ \citep[e.g.,][]{2003MNRAS.339..793G, 2010ApJ...724..559L, 2012MNRAS.419.2095M}.
Furthermore, the dispersion of $L_{\rm X}/{\rm SFR}$ is at most $\sim 0.4$ dex \citep{2012MNRAS.419.2095M}.
Assuming a simple power-law spectrum with $L_{\rm X}(\nu) \propto \nu^{-\alpha}$ 
(see equation~\ref{eq:spect}),
we can estimate the X-ray flux at $1$ keV 
(in units of $10^{-21}~{\rm erg}~{\rm s}^{-1}
{\rm cm}^{-2}~{\rm sr}^{-1}~{\rm Hz}^{-1}$) as
\begin{eqnarray}
J_{\rm X,21}=
\left\{
\begin{array}{c}
0.68\\
0.89\\
1.2
\end{array}
\right\}
\times 4\times 10^{-4}\left(\frac{{\rm SFR}}{\msunyr}\right)
\left(\frac{d}{10~{\rm kpc}}\right)^{-2},
\label{eq:x_corr}
\end{eqnarray}
where the three values in brackets correspond to cases for $\alpha=1.6$, $1.8$, and $2.0$
(from top to bottom).

Several studies have investigated the redshift evolution of the $L_{\rm X}$-SFR relation 
using empirical data.
The linear relation observed in local star-forming galaxies ($z=0$)
does not change significantly up to $z\la 2$ 
\citep{2003MNRAS.339..793G, 2008ApJ...681.1163L, 2014MNRAS.437.1698M}.
The Chandra Deep Field-South suggests that the ratio increase as $\propto (1+z)$ 
out to $z\sim 4$ \citep{2013ApJ...762...45B}.
Furthermore, the existence of the unresolved soft X-ray background places
a constraint on its evolution at higher redshifts:
${\rm d}\log (L_{\rm X}/{\rm SFR})/{\rm d}\log (1+z) \leq 1.3$
\citep{Dijkstra+12}.

As noted above, the latest simulations suggest that
Pop III stars tend to form with large ($>10~\msun$) masses 
in binary or multiple systems.
Thus, we expect more X-ray binaries in the high-$z$ Universe than in the local galaxies.
Although the properties of Pop III binaries remain highly uncertain, 
\cite{2014arXiv1407.1847H} estimate that they produced an X-ray background intensity
$J_{\rm X,21}\sim 0.03$ at $z\sim 20$.
This value is a few hundred times larger than the $L_{\rm X}/{\rm SFR}$ of low-$z$ galaxies.
Similarly, population synthesis models of \cite{2013ApJ...764...41F,2013ApJ...776L..31F}
predict that  $L_{\rm X}/{\rm SFR}$ at $z\sim 10$ is higher than the local value by an order of magnitude.

To keep our results and discussions conservative,
we here adopt the X-ray intensity using the $L_{\rm X}$-SFR 
relation obtained from observations of low-$z$ galaxies 
(instead of theoretical extrapolations of the ratio to higher redshifts).
We define the dimensionless number in curly brackets
in equation (\ref{eq:x_corr}) as $f_{\rm X}\equiv J_{\rm X,21}/(4\times 10^{-4})\simeq 1$
and treat it as a parameter set fiducially to unity
(note that $f_{\rm X}$ has an empirical dispersion of 
$0.4$ dex in low-z galaxies; \citealt{2012MNRAS.419.2095M}).
As we describe above, both observations and theoretical works 
suggest $f_{\rm X}\ga 1-10$.
We note that our fiducial model is in close agreement with \cite{2013MNRAS.431..621M}, 
who used/found $f_{\rm X}\approx 1$ based on
the number of X-ray photons per stellar baryon $N_{\rm X}\approx 0.2$ and 
the fraction of baryons converted into stars $f_\ast \approx 0.1$
\footnote{While we choose $f_{\rm X}=1$ as our fiducial model, the fiducial models of most previous
theoretical studies \citep[e.g.][]{2006MNRAS.371..867F,2007MNRAS.376.1680P}
correspond to $f_{\rm X}\approx 4-5$ in our notation.}.

\subsubsection{FUV radiation}
\label{seq:FUV_rad}
Next, we estimate the LW intensities from star-forming galaxies
consisting of Pop II ($Z=10^{-3}$) and Pop III stars ($Z=0$).
We adopt the Salpeter initial mass function ($1\leq M_\ast \leq 100~\msun$).
The number flux of LW photons is estimated as 
$Q_{\rm LW}=5.25~(3.72)\times 10^{53}~{\rm s}^{-1}~({\rm SFR}/\msunyr)$
for the Pop II (III) case, assuming constant star formation \citep{2003A&A...397..527S}.
To estimate the mean intensity at $13.6$ eV, we here consider 
two types of spectral models of star-forming galaxies:
(i) a thermal spectrum with the effective temperature of $\geq 2\times 10^4$ K
and (ii) a flat spectrum ($\propto \nu ^{0-0.5}$ at $1\la h\nu \leq 13.6$ eV), 
which may be expected because of the superposition of radiation 
from low-mass and massive stars \citep{2011MNRAS.415.2920I}.
We find
\begin{eqnarray}
J_{\rm LW,21}=
\left\{
\begin{array}{c}
0.85\\
1.1\\
1.0\\
1.3
\end{array}
\right\}
\times 90\left(\frac{{\rm SFR}}{\msunyr}\right)
\left(\frac{d}{10~{\rm kpc}}\right)^{-2},
\label{eq:LW_corr}
\end{eqnarray}
where the first three values in brackets correspond to cases for thermal spectra
with $T_\ast = 2\times 10^4$, $3\times 10^4$ and $10^5$ K 
(with the third value corresponding to the PopIII case),
and the fourth value to the flat spectrum.
The actual effective temperatures of Pop II galaxies are hotter than $10^4$ K 
\citep{2011MNRAS.415.2920I}.
The ratio of $k_{\rm H^-}^{\rm pd}/k_{\rm H_2}^{\rm pd}$, which is a good indicator of $J_{\rm crit,0}$ 
\citep{2014MNRAS.445..544S}, decreases for higher $T_\ast$.
We summarize the the values of these ratios for spectral models
of Pop II galaxies in Table \ref{tab_galsp}.
For most cases (except for the instantaneous starburst model with $2\times 10^{-2}~\zsun$), the ratio is smaller than $60$.
Thus, as far as the critical FUV intensity is concerned, 
these spectral models are closest to the case of
a thermal spectrum with $T_\ast \simeq 3\times 10^4$ K (see Table \ref{tab_phch}).
We will call the dimensionless factor in the curly brackets as $f_{\rm LW}(\simeq 1)$.

\begin{table}
\begin{center}
\caption{The ratio of the photodissociation rates of H$_2$ and H$^-$ for Pop II galaxies.}
\begin{tabular}{c}
constant star formation: $100$ Myr$^a$
\end{tabular}
  \begin{tabular}{c|c|c|c}\hline
Z($\zsun$)& $0$ & $5\times10^{-4}$ & $2\times 10^{-2}$\\
$k_{\rm H^-}^{\rm pd}/k_{\rm H_2}^{\rm pd}$ & $4.2\times 10$& $3.9\times 10$ & 
$5.4\times 10$ 
\end{tabular}
\vspace{4mm}
\begin{tabular}{c}
instantaneous starburst : $100$ Myr$^b$
\end{tabular}
  \begin{tabular}{c|c|c|c}\hline
Z($\zsun$)& $0$ & $5\times10^{-4}$ & $2\times 10^{-2}$\\
$k_{\rm H^{-}}^{\rm pd}/k_{\rm H_2}^{\rm pd}$ & $3.8\times 10$& $5.6\times 10$ & 
$2.2\times 10^2$ \\\hline
\end{tabular}
   \label{tab_galsp}
  \end{center}
$^a$The duration of star formation.\\
$^b$The time since the starburst.\\
References: \cite{2011MNRAS.415.2920I, 2014MNRAS.445..544S}.
\end{table}

\subsubsection{Relation between X-rays and FUV radiation}
Combining the expressions for $J_{\rm X}$ and $J_{\rm LW}$ above,
we obtain
\begin{equation}
\frac{J_{\rm X,21}}{ J_{\rm LW,21}}
\simeq 4.4\times 10^{-6}~\left(\frac{f_{\rm X}}{f_{\rm LW}}\right).
\label{eq:uv_xray}
\end{equation}
Above, $f_{\rm X}\ga 1-10$ is the normalization
of the X-ray intensity with respect to low-$z$ star-forming galaxies 
(\S\ref{sec:Xray-flux}, eq. \ref{eq:x_corr}),
and $f_{\rm LW}\simeq 1$ is the dependence of the $J_{\rm LW}$ normalization
on the galaxy FUV spectrum (\S\ref{seq:FUV_rad}, eq. \ref{eq:LW_corr}).
For example, if a star-forming galaxy has
the same ratio of X-rays to FUV photons as the value typically found in
lower-redshift galaxies ($ f_{\rm X}/f_{\rm LW}=1$)
and irradiates a neighboring halo at an FUV intensity $J_{\rm LW,21}\ga 2\times 10^3$,
then it will simultaneously expose it to a soft X-ray intensity $J_{\rm X,21}\ga 0.01$.
In what follows, we consider a wide range $0.1\leq f_{\rm X}/f_{\rm LW}\leq 10$ 
in order to present a conservative discussion.

\subsection{Critical LW intensity}
\label{subsec:actual_jcr}
\begin{figure}
\epsfig{file=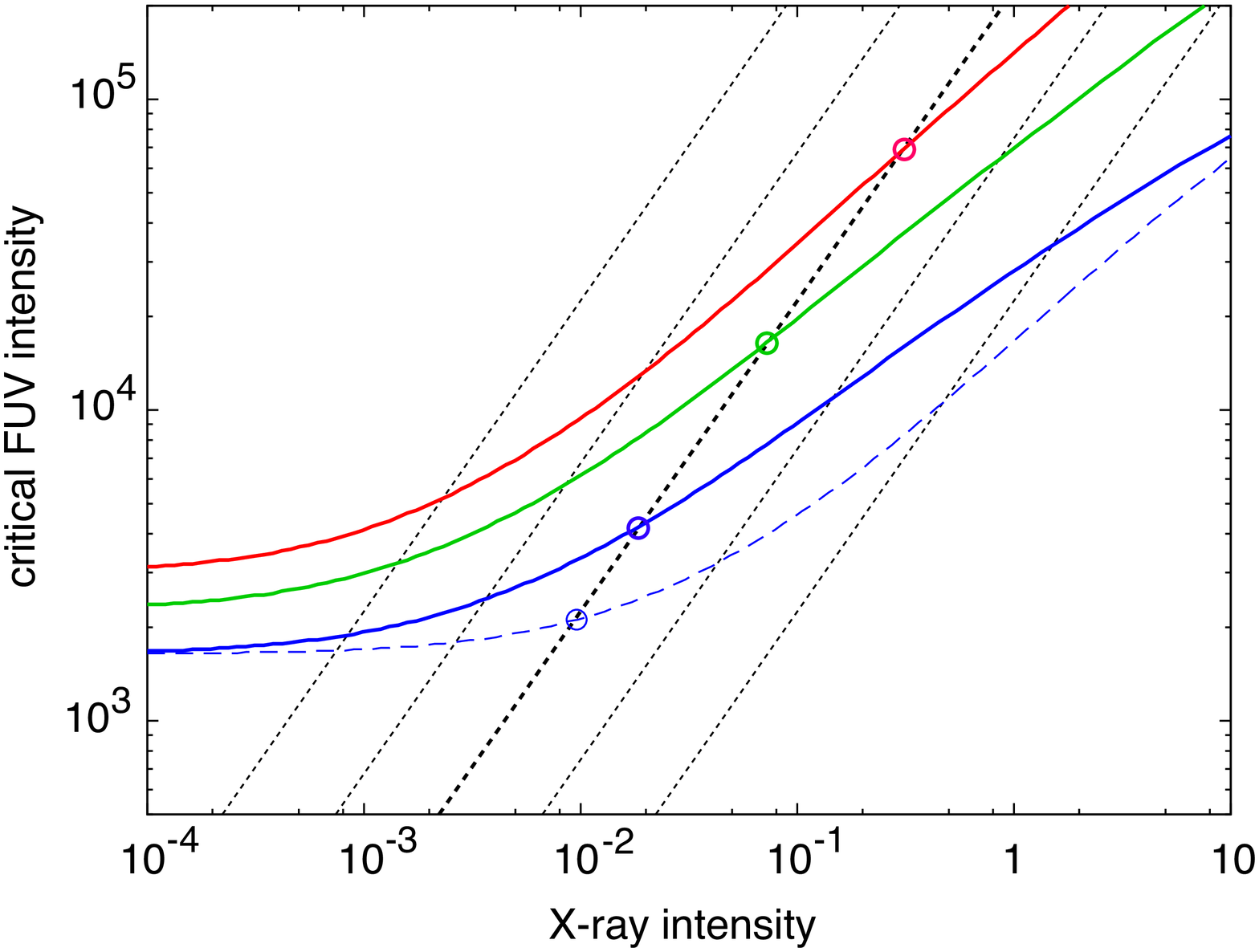,height=60mm,width=80mm}
\caption{
$J_{\rm crit}$ vs. $J_{\rm X}$.
The critical LW intensity $J_{\rm crit}$ as a function 
of the X-ray intensity $J_{\rm X,21}$ for $T_\ast =2\times 10^4$, $3\times 10^4$ and $10^5$ K from bottom to top
(solid curves; $h\nu_{\rm min}=1$ keV) and for $T_\ast =2\times 10^4$ K 
(dashed curve; $h\nu_{\rm min}=2$ keV).
Dotted lines show the relashion between the LW and X-ray intensities
given by equation (\ref{eq:uv_xray}):
fiducial model ($f_{\rm X}/f_{\rm LW}=1$; thick dotted) and the cases for
$f_{\rm X}/f_{\rm LW}=1/10$, $1/3$, $3$ and $10$ from left to right (thin dotted).
A colour version of this figure is available in the online version.}
\label{fig:2}
\end{figure}

In Fig.~\ref{fig:2}, we show how the critical LW intensity $J_{\rm crit}$ 
increases when accounting for X-ray ionizations.
This figure shows the relationship between the X-ray-corrected value ($J_{\rm crit}$) 
and the value calculated assuming a UV-only spectrum ($J_{\rm crit,0}$).
As described above, the critical LW intensity $J_{\rm crit}$ increases
in the presence of an X-ray flux. 
For the purposes of computing $J_{\rm crit}$,
the spectra of Pop II galaxies are well
described by a thermal spectrum with $T_\ast \simeq 3\times 10^4$ K  (see \S\ref{seq:FUV_rad}).
In what follows, we consider this case as our fiducial model.
For $h\nu_{\rm min}=1$ keV (solid curves), the X-ray-corrected
value can be fit by the following simple formula (\S\ref{sec:method}):
\begin{equation}
J_{\rm crit}=J_{\rm crit,0}\left(1+\frac{J_{\rm X,21}}{2.2\times 10^{-3}}\right)^{0.56}.
\label{eq:J_cr}
\end{equation}
where $J_{\rm crit,0}$ is the value calculated without considering 
X-ray ionizations at all.
Using equation~(\ref{eq:uv_xray}), the actual critical LW intensity $J_{\rm crit}$ can be 
written in terms of the original critical LW intensity (i.e., no X-ray flux) as
\begin{equation}
J_{\rm crit}\simeq 1.8\times 10^4\left(\frac{J_{\rm crit,0}}{2.4\times 10^3}\right)^{2.3}
\left(\frac{f_{\rm X}}{f_{\rm LW}}\right)^{1.3},
\label{eq:J_cr2}
\end{equation}
which corresponds to an intersection of the solid curve (green) and thick dotted line in Fig.~\ref{fig:2}.
This equation is approximately valid for 
$f_{\rm X}/f_{\rm LW}\ga 0.2~(J_{\rm crit,0}/2.4\times 10^3)^{-1}$;
at these values, X-ray ionization suppresses DCBH formation by raising the value
of the critical flux $J_{\rm crit}$ necessary to keep $\Hmol$ photodissociated.

In Table.~\ref{tab_JJ}, we summary the values of $J_{\rm crit}/J_{\rm crit,0}$
at the intersection points between solid curves and dashed lines in Fig.~\ref{fig:2} 
for various cases.
If $f_{\rm X}/f_{\rm LW}\sim 1$, and if the X-ray-uncorrected value $J_{\rm crit,0} \ga 2\times 10^3$,
then soft X-rays would increase $J_{\rm crit}$ by a factor of $\sim 3-10$.
As we will discuss in the next section, even such a modest increase in 
$J_{\rm crit}$  is expected to decrease the abundance of potential DCBH sites by several orders of magnitude
(\citetalias{Dijkstra+14}).
However, we have earlier described calculations showing that $f_{\rm LW}\sim 1$ for several
irradiation spectra, and $f_{\rm X}\sim 1-10$ for $z\ga 10$ galaxies.
Taken together, those results suggest $f_{\rm X}/f_{\rm LW} > 1$ at redshifts
relevant for direct collapse, in which case
the suppression of DCBH sites is much more severe 
than in our fiducial, conservative case $f_{\rm X}/f_{\rm LW}=1$.

\begin{table}
\begin{center}
\caption{The enhancement factor of the critical FUV intensity ($J_{\rm crit}/J_{\rm crit,0}$)
at the intersection points between solid curves and dashed lines in Fig.~\ref{fig:2}.
The first three columns are the cases for $h\nu_{\rm min}=1$ keV and the fourth is for $h\nu_{\rm min}=2$ keV.}
  \begin{tabular}{c|c|c|c|c}\hline
$T_\ast$ (K) & $2\times 10^4$ & $3\times 10^4$ & $10^5$ & $(2\times 10^4$; 2 keV)\\\hline
$f_{\rm X}/f_{\rm LW}=1$ & 2.5 & 7.0 & 22 & 1.3\\
$f_{\rm X}/f_{\rm LW}=3$ & 6.5 & 26 & 120 & 1.9\\\hline
\end{tabular}
   \label{tab_JJ}
  \end{center}
\end{table}

\begin{table}
\begin{center}
\caption{Summary of the critical LW intensity $J_{\rm crit,0}$}
  \begin{tabular}{l|c|c|c} \hline
authors & $J_{\rm crit,0}$ & $T_\ast$ (K) & method \\ \hline\hline
SBH10& $39$ & $10^4$ & one-zone\\
 & $30-300$ & $10^4$ & 3D \\
IO11 & $20$ & $10^4$ & one-zone\\
L14 & $400-1500$ & $10^4$ & 3D\\
SOI14 & $25$ & $10^4$ & one-zone\\
\hline
SOI14& $\ga 1400$ & $>2\times 10^4$ & one-zone\\
\hline
SBH10&$1.2\times 10^4$& $10^5$&one-zone\\
 &$10^4-10^5$& $10^5$&3D\\
IO11  & $1.6\times 10^4$ & $10^5$ & one-zone\\
WHB11 & $2-4\times 10^3$& $10^5$ & 3D \\\hline
   \end{tabular}
   \label{tab_jcr}
  \end{center}
References: 
\cite{2010MNRAS.402.1249S} (SBH10);
\cite{IO11} (IO11);
\cite{2011MNRAS.418..838W} (WHB11);
\cite{2014MNRAS.443.1979L} (L14);
\cite{2014MNRAS.445..544S} (SOI14).
\end{table}

Equation~(\ref{eq:J_cr2}) shows that $J_{\rm crit}$ is sensitive to the value of $J_{\rm crit,0}$. 
The higher the value of $J_{\rm crit,0}$ (e.g. higher $T_\ast$), the critical LW flux evaluated 
without taking into account X-ray ionizations, the higher the value of the effective value $J_{\rm crit}$. 
The relative enhancement is roughly proportional to the FUV intensity itself.
In addition, we find that soft X-rays are suitable for increasing $J_{\rm crit}$ more than hard X-rays
because $f_{\rm X}$ is effectively smaller by one oder of magnitude for the case of hard X-rays.
For $h\nu_{\rm min}=2$ keV (dashed line in Fig.~\ref{fig:2}), the X-ray-corrected value is 
almost the same as the original value.

The critical LW intensity evaluated without consideration of external ionizations, 
$J_{\rm crit,0}$, has been investigated using a one-zone model 
\citep[e.g.,][]{O01,IO11}
and three-dimensional numerical simulations 
\citep[e.g.,][]{2003ApJ...596...34B,2010MNRAS.402.1249S,2011MNRAS.418..838W,
2014MNRAS.443.1979L}.
We summarize the results of these previous studies in Table~\ref{tab_jcr}.
For a soft spectrum with $T_\ast=10^4$ K, the values of $J_{\rm crit,0}$ are 
$20-40$ (one-zone models)
and $30-10^3$ (3D simulations).
For harder spectra with $T_\ast >2\times 10^4$ K, the values of $J_{\rm crit,0}$ are 
$\ga 10^3$ (one-zone models) and $\sim 10^4$ (3D simulations).
\cite{2014MNRAS.445..544S} found that $J_{\rm crit,0}$ does not change significantly
for $T_\ast \ga 2\times 10^4$ K because the typical Pop II galaxies have spectra 
that are flatter and harder than the thermal spectrum with $T_\ast =10^4$ K \citep{2011MNRAS.415.2920I}.
The values of $J_{\rm crit,0}$ estimated from the 3D simulations tend
to be larger by one order of magnitude than that derived in one-zone models
because of $\sim 20\%$ spatial variation in the temperature 
inside the collapsing gas clouds.
The temperature fluctuations produce a large difference in the H$_2$-collisional 
dissociation rate by one order of magnitude \citep{2010MNRAS.402.1249S}.
In light of this effect, it is reasonable to expect the actual value of $J_{\rm crit,0}$ to be $\ga 10^3$.

The value of $J_{\rm crit,0}$ has several uncertainties
beyond the FUV spectrum of the irradiating galaxies.
For example, we can list uncertainties from (i) the H$_2$ self-shielding factor,
as well as (ii) reaction rate coefficients associated with H$_2$ and H$^-$ formation
\citep[see also][]{2015arXiv150105960G}. 
These can affect the value of $J_{\rm crit,0}$ by a factor of $\sim 3-4$.
Therefore, in what follows,
we do not specify the value of $J_{\rm crit,0}$ but regard as a free parameter,
and choose instead to focus on the relative enhancement $J_{\rm crit}/J_{\rm crit,0}$.

\section{DCBH formation probability and number density}
\label{sec:prob}

\begin{figure}
\epsfig{file=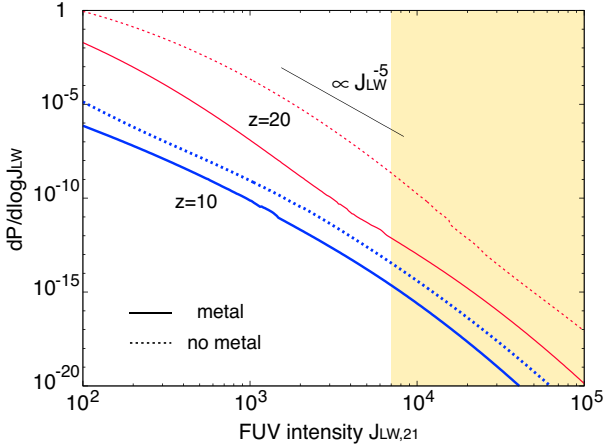,height=60mm,width=80mm}
\caption{The probability distribution function (PDF) of atomic-cooling halos (mass $M_{\rm ac}$) 
being exposed to an FUV intensity $J_{\rm LW}$ by a neighbouring star-forming galaxy. 
The thick (blue) lines show the cases at $z=10$, and the thin (red) lines show the cases at $z=20$. 
The solid and dashed lines show the cases with with and without metal enrichment, respectively, 
by galactic outflows from the FUV source. 
The PDF is well-characterised by a power law with index $\approx 5$ 
in the range $J_{\rm LW,21}\sim 10^3-10^4$. 
The shaded region shows the required $J_{\rm LW}$ intensity for forming DCBHs ($J_{\rm crit}$), 
if $J_{\rm crit,0} =1500$ and $f_{\rm X}/f_{\rm LW}=1$. 
A colour version of this figure is available in the online version.}
\label{fig:PDF}
\end{figure}

\begin{table}
\begin{center}
\caption{Summary of how the critical FUV intensity for DCBH formation ($J_{\rm crit}$) 
and the corresponding formation probability ($P_{\rm DCBH}$) increase 
when accounting for X-ray ionization. 
We show how these quantities change for three representative values of 
the X-ray-{\it uncorrected} value $J_{\rm crit,0}$ ($500$, $1500$ and $3000$); 
for $z=10$ and $z=20$; with and without metal enrichment by galactic winds 
(\citetalias{Dijkstra+14} model). 
These calculations assume $f_{\rm X}/f_{\rm LW}=1$.}
  \begin{tabular}{c|c|c|c|c} \hline
  & & UV only &&UV + X\\ \hline \hline
  & & $J_{\rm crit,0}=5\times 10^2$ & $\Rightarrow$ & $J_{\rm crit}\simeq 9\times 10^2$ \vspace{1mm}\\
$z=10$ & w/wind & $1.6\times 10^{-10}$ &  & $1.2\times 10^{-11}$\\
             & no wind & $1.6\times 10^{-9}$ &  & $1.4\times 10^{-10}$\\ 
$z=20$ &w/wind& $5.9\times 10^{-7}$ &  & $1.8\times 10^{-8}$\\
            &no wind&  $4.9\times 10^{-4}$ &  & $3.4\times 10^{-5}$\\ \hline
             & & $J_{\rm crit,0}=1.5\times 10^3$ & $\Rightarrow$ & $J_{\rm crit}\simeq 7\times 10^3$ \vspace{1mm}\\
$z=10$ & w/wind & $7.2\times 10^{-13}$ &  & $2.0\times 10^{-16}$ \\
             & no wind &  $1.2\times 10^{-11}$ &  &  $3.1\times 10^{-15}$\\ 
$z=20$ &w/wind& $6.0\times 10^{-10}$ &  & $7.1\times 10^{-14}$ \\
            &no wind&  $2.1\times 10^{-6}$ &  & $1.7\times 10^{-10}$ \\ \hline
             & & $J_{\rm crit,0}=3\times 10^3$ & $\Rightarrow$ & $J_{\rm crit}\simeq 3\times 10^4$ \vspace{1mm}\\
$z=10$ & w/wind & $2.1\times 10^{-14}$ &  & $5.0\times 10^{-21}$ \\
             & no wind &  $3.4\times 10^{-13}$ &  &  $1.2\times 10^{-19}$\\ 
$z=20$ &w/wind& $7.8\times 10^{-12}$ &  & $5.6\times 10^{-18}$ \\
            &no wind&  $3.4\times 10^{-8}$ &  & $1.6\times 10^{-15}$ \\ \hline

   \end{tabular}
   \label{tab_jcr_jcr0}
  \end{center}
\end{table}

Now, we turn to the probability that a massive halo is exposed to $J_{\rm crit}$ 
by a neighbouring galaxy, and use this quantity to estimate the number density 
of $z\sim 10$ seed BHs formed through FUV-aided direct collapse. 
The methods and calculations presented here follow those developed by 
\citetalias{Dijkstra+14}. 
We summarize the requisite calculations below, and refer the reader to that paper 
for further details.

Metal pollution by galactic outflows could suppress the probability of DCBH formation 
because the metal cooling induces efficient gas fragmentation.
We here consider the case incorporating the metal-enriching wind model of \citetalias{Dijkstra+14},
as well as the case without winds.

We briefly describe how to calculate the PDF of $J_{\rm LW}$.
We define potential DCBH formation sites as atomic-cooling halos
with virial temperature of $T_{\rm vir}=10^4$ K,
corresponding to a mass $M_{\rm ac}(z)=8.1\times 10^7~\msun~((1+z)/11)^{-3/2}$.
A nearby star-forming galaxy can act
as an FUV source for keeping $\Hmol$ photodissociated \citep{Dijkstra+08}.
The differential probability distribution of finding an FUV source with mass $M$ 
at a distance $r$ is simply written
\begin{equation}
\frac{{\rm d}^2P_1(M,r,z)}{{\rm d}M~{\rm d}r}=4\pi r^2 (1+z)^3
\left[1+\xi(M_{\rm ac},M,r,z)\right]\frac{{\rm d}n_{\rm ST}}{{\rm d}M},
\end{equation}
where $\xi$ is the non-linear bias function, which represents the clustering of the two halos
(i.e. the excess probability of finding another halo at a distance $r$; 
\citealt{2003MNRAS.341...81I})
and ${\rm d}n_{\rm ST}/{\rm d}M$ is the Sheth-Tormen halo mass function
\citep{2001MNRAS.323....1S}.
Furthermore, we assume a log-normal distribution for the distribution of the LW luminosity 
of the source galaxies,
\begin{align}
\frac{{\rm d} P_2(L_{\rm LW},M,z)}{{\rm d}\log L_{\rm LW}} &= \frac{1}{\sqrt{2\pi \sigma_{\rm LW}^2}}\nonumber\\
\times &\exp \left[-\frac{(\log L_{\rm LW} - \log \langle L_{\rm LW}\rangle)^2}
{2\sigma_{\rm LW}^2}\right],
\end{align}
where $\langle L_{\rm LW}\rangle$ is the mean LW luminosity (see equation 6 and 8 in \citetalias{Dijkstra+14}) and 
$\sigma_{\rm LW}=0.4$ is the dispersion.
While our primary motivation in choosing a log-normal distribution
for $L_{\rm LW}$ is consistency with \cite{Dijkstra+08} and \citetalias{Dijkstra+14},
we also note that this type of distribution is commonly used to describe populations of galaxies
\citep[e.g.][]{2005ApJ...627L..89C, 2008MNRAS.383..355V}.
Using the relation $L_{\rm LW}=16\pi^2 r^2 J_{\rm LW}$, we obtain the PDF
of a given atomic-cooling halo being exposed to a LW flux $J_{\rm LW}$:
\begin{equation}
\frac{{\rm d}P_{\rm DCBH}(J_{\rm LW},z)}{{\rm d}\log J_{\rm LW}}=\int _{M_{\rm min}}^\infty {\rm d}M 
\int_{r_{\rm min}}^\infty {\rm d}r\frac{{\rm d}^2P_1}{{\rm d}M~{\rm d}r} \frac{{\rm d}P_2}{{\rm d}\log L_{\rm LW}}.
\label{eq:PDF}
\end{equation}
We set $M_{\rm min}=M_{\rm ac}$.
For the case with metal pollution by galactic winds, we apply
$r_{\rm min}={\rm max}(r_{\rm vir}(M_{\rm ac})+r_{\rm vir}(M), r_{s}(M))$,
where $r_{s}(M)$ is the distance from the sources within which the gas is polluted by metals and
DCBH formation is quenched (see equation 5 in \citetalias{Dijkstra+14}).
For the case without metal-enriching winds, 
we simply set $r_{\rm min}=r_{\rm vir}(M_{\rm ac})+r_{\rm vir}(M)$.
Note that \citetalias{Dijkstra+14} set $r_{\rm min}=2r_{\rm vir}(M_{\rm ac})$.

In Fig.~\ref{fig:PDF}, we show the PDF of $J_{\rm LW}$ for two redshifts 
of $z=10$ (blue) and $20$ (red).
The solid and dashed lines represent the cases with and without metal 
pollution by galactic outflows, respectively. 
With (without) metal pollution\footnote{Both values are smaller than the results 
shown in Fig. C1 of \citetalias{Dijkstra+14}. 
We have determined that this difference is due to our choice of 
$r_{\rm min} = r_{\rm vir}(M_{\rm ac})+r_{\rm vir}(M)$, which is greater than the choice 
$r_{\rm min}=2r_{\rm vir}(M_{\rm ac})$ adopted by \citetalias{Dijkstra+14}.
Larger $r_{\rm min}$ results in lower probabilities of being exposed to a given $J_{\rm LW}$,
since the requisite luminosity $L_{\rm LW}\propto r^2 J_{\rm LW}$ follows a log-normal PDF.
Note that this also means that ${\rm d}P_{\rm DCBH}/{\rm d}\log J_{\rm LW}$ 
decreases with redshift
since $r_{\rm vir}(M_{\rm ac})\propto (1+z)^{-3/2}$
and $r_{\rm vir}(M)\propto M^{1/3}(1+z)^{-1}$.}, 
${\rm d}P_{\rm DCBH}/{\rm d}\log J_{\rm LW} \sim 8\times10^{-11}
(9\times 10^{-10})$ for $J_{\rm LW}=10^3$ at $z\simeq 10$.
The PDF rapidly falls off with increasing $J_{\rm LW}$.
For $10^3 \la J_{\rm LW,21} \la 10^4$, this behavior roughly follows
a power law $\propto J_{\rm LW}^{-\beta}$ with $\beta \approx 5$.

\begin{figure}
\epsfig{file=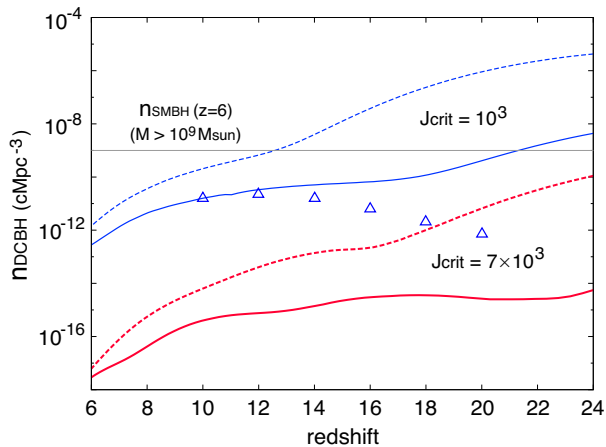,height=60mm,width=80mm}
\caption{Redshift evolution of the number density (comoving) of forming 
DCBHs in atomic cooling halos.
The thin (blue) and thick (red) lines show the cases that the X-ray-corrected 
$J_{\rm crit}=10^3$ and $7\times 10^3$, respectively.
The solid and dashed lines represent the cases with and without metal-enrichment
by galactic winds, respectively.
The horizontal line indicates $10^{-9}$ cMpc$^{-3}$, which is the number density
of observed SMBHs with mass $\ga 10^9~\msun$ at $z\ga 6$.
Triangle symbols show the results for $J_{\rm crit}=10^3$ and with galactic winds,
which are calculated using the same treatment of the nonlinear bias function as \citetalias{Dijkstra+14}.
A colour version of this figure is available in the online version.}
\label{fig:n_DCBH}
\end{figure}

The steep dependence of ${\rm d}P_{\rm DCBH}/{\rm d}\log J_{\rm LW}$ has a dramatic effect on 
the probability of DCBH formation. 
Using the approximation above that ${\rm d}P_{\rm DCBH}/{\rm d}\log J_{\rm LW}
\propto J_{\rm LW}^{-\beta}$ with $\beta\approx 5$, and combining this with equation (\ref{eq:J_cr2}), 
we can write how the X-ray-corrected PDF depends on $J_{\rm crit,0}$:
\begin{align}
\frac{{\rm d}P_{\rm DCBH}(z=10)}{{\rm d}\log J_{\rm LW}}\approx 
\left\{
\begin{array}{l}
2\times 10^{-15}\\
3\times 10^{-14}
\end{array}
\right\}
\times \left(\frac{J_{\rm crit,0}}{1500}\right)^{-12}
\left(\frac{f_{\rm X}}{f_{\rm LW}}\right)^{-6.7},
\end{align}
where the two values inside the curly brackets correspond to cases 
with (top) and without (bottom) metal-enrichment. 
In Table \ref{tab_jcr_jcr0}, we summarize the values of the integrated probability 
$P_{\rm DCBH} (\geq J_{\rm crit},z)$ for $z=10$ and $z=20$. 
As we showed in \S\ref{subsec:actual_jcr}, X-ray ionizations can increase $J_{\rm crit}$ 
if the X-ray-{\it uncorrected} value is 
$J_{\rm crit,0}\ga 500~(f_{\rm X}/f_{\rm LW})^{-1}$.
This decreases $P_{\rm DCBH}$ by several orders of magnitude.

Above, we showed that soft X-rays can increase the critical FUV intensity
to $J_{\rm crit} \sim 7\times 10^3$ if $J_{\rm crit,0}$ (the value without
accounting for X-ray ionizations) is $1.5\times 10^3$. 
This results in a DCBH formation probability (per halo): 
$P_{\rm DCBH} \sim 2.0\times 10^{-16}~(3.1\times 10^{-15})$
at $z\sim10$ with (without) metal enrichment by galactic winds.
(Note that we have not accounted for enrichment via \textit{in situ} star formation,
which would further reduce $P_{\rm DCBH}$.)
These values are $3-4$ orders of magnitude lower than the result 
calculated without considering soft X-rays:  
$P_{\rm DCBH} \sim 7.2\times 10^{-13}~(1.2\times 10^{-11})$.
This case corresponds to an X-ray intensity 
$J_{\rm X,21}\simeq 0.07$
that is larger by a few orders of magnitude than that of the observed X-ray background.
This is only natural, as to be exposed to FUV intensities above $J_{\rm crit}$,
the putative DCBH formation site must be very close to a star-forming galaxy,
and therefore also exposed to strong soft X-ray intensities.

Finally, the number density (comoving) of forming DCBHs in an atomic cooling halo 
with mass $M_{\rm ac}$
is given by
\begin{align}
n_{\rm DCBH}(z)&=\int_{M_{\rm ac}}^\infty {\rm d}M
\frac{{\rm d}n_{\rm ST}}{{\rm d}M}P_{\rm DCBH}(\geq J_{\rm crit},z),\nonumber\\
&\approx n_{\rm ST}(M\geq M_{\rm ac},z)~P_{\rm DCBH}(\geq J_{\rm crit},z).
\label{eq:nDCBH}
\end{align}
Here, we have followed the approximations taken by \citetalias{Dijkstra+14},
except that we have taken the factor $P_{\rm gen}$ (the probability that a given halo has {\it not}
been metal-enriched by {\it in situ} star formation) to be unity for simplicity.
Note that since $P_{\rm gen}\le 1$, the estimate given by
equation (\ref{eq:nDCBH}) is conservative.

In Fig.~\ref{fig:n_DCBH}, we show $n_{\rm DCBH}$ as a function of $z$
for $J_{\rm crit}=10^3$ (thick red) and $7\times 10^3$ (thin blue).
The solid and dashed lines represent the cases with and without metal-enrichment
by galactic winds, respectively.
At $z \simeq 10$, the number density of DCBHs is smaller than that of $\ga 10^9~\msun$ 
SMBHs observed at $z\ga 6$ ($\sim 10^{-9}$ cMpc$^{-3}$, \citealt{2010AJ....139..906W}; 
shown as a horizontal line in the figure), for all cases considered here.
For $J_{\rm crit}=7\times 10^3$, which corresponds to $J_{\rm crit,0}\simeq 1.5\times 10^3$
(no X-ray ionization) for $f_{\rm X}/f_{\rm LW}=1$, 
even the case without metal enrichment cannot exceed
$10^{-9}$ cMpc$^{-3}$ for all redshift.
We note that $n_{\rm DCBH}$ for the case of $J_{\rm crit}=10^3$ 
and no metal pollution can be larger than the observed value at $z\ga 12$
although this is the most optimistic case (i.e. no winds and $P_{\rm gen}=1$).

\begin{figure}
\epsfig{file=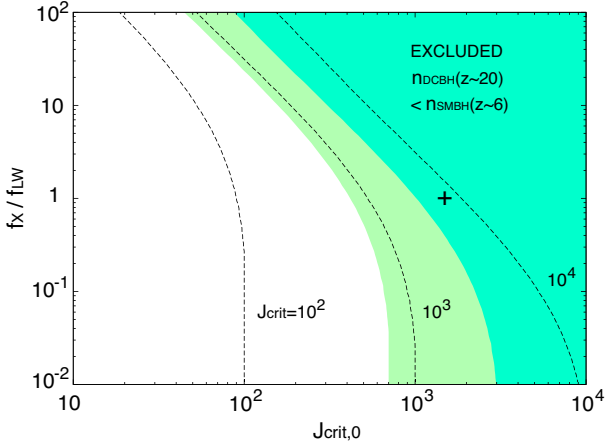,height=60mm,width=80mm}
\caption{Contour plots of $J_{\rm crit}$ (including X-ray ionization) 
in the two-dimensional 
parameter space ($J_{\rm crit,0}$, $f_{\rm X}/f_{\rm LW}$).
The shaded regions show the regions in this parameter space where the number density 
of DCBHs falls below the value of $\sim 10^9~\msun$ SMBHs observed at $z\sim 6$.
The lightly (darkly) shaded region shows this space for $n_{\rm DCBH}$ 
evaluated at $z=10~(20)$, 
corresponding to $J_{\rm crit} \ga 700~(3\times 10^3)$. 
Cross symbol indicates our fiducial model ($J_{\rm crit,0}=1.5\times 10^3$ and 
$f_{\rm X}/f_{\rm LW}=1$).
A colour version of this figure is available in the online version.}
\label{fig:5}
\end{figure}

Note that we find $n_{\rm DCBH}$ that increases toward higher redshifts,
whereas it decreases in similar calculations by \citetalias{Dijkstra+14} (their Fig. 4).
This is due to the fact that  \citetalias{Dijkstra+14} computed
$\xi(z=10)$, the two-point correlation function at $z=10$,
and assumed that it can be approximated at any other redshift as
$\xi\approx\xi(z=10)[D_+(z)/D_+(z=10)]^2$, where $D_+(z)$ is the growth factor.
This assumption that $\xi\propto D_+^2$ is valid for the linear growth factor;
however, for the nonlinear growth factor formulae of \cite{2003MNRAS.341...81I}
at small separations, $\xi$ does not follow this relation and instead increases with $z$.
This leads to the qualitative difference in our findings.
For reference, in Fig.~\ref{fig:n_DCBH} we have overlaid with triangle symbols
the $n_{\rm DCBH}$ estimates calculated with the same bias treatment as \citetalias{Dijkstra+14}.
Again: our calculations for $n_{\rm DCBH}$ increases with $z$, because we explicitly
calculate the nonlinear bias at each redshift using the equations of  \cite{2003MNRAS.341...81I};
in contrast,  \citetalias{Dijkstra+14} assumed  $\xi(z)\propto D_+^2(z)$,
which results in  $n_{\rm DCBH}$ decreasing toward higher redshifts.

Note that for combinations of high $J_{\rm crit}$ and lower redshifts
(e.g. $J_{\rm crit}\ga 10^3$ and $z\la 15$),
the characteristic halo mass where $P_{\rm DCBH}$
is highest can increase to $M\gg M_{\rm ac}$.
This is because $r_{\rm min}$ increases toward low $z$ as described above.
Since the required LW luminosity for a neighbouring halo to enable direct collapse
is $L_{\rm crit}>16\pi^2 r_{\rm min}^2 J_{\rm crit}$,
for cases of large $J_{\rm crit}$ and large $r_{\rm min}$ (low $z$)
only massive neighbouring halos are able to provide such a flux.
In such cases, the probability is limited not by the number of
atomic-cooling halos, but by the number of massive FUV sources,
and equation (\ref{eq:nDCBH}) should be evaluated using the abundances
of the latter.

In Fig.~\ref{fig:5}, we show contour plots of $J_{\rm crit}$ (including X-ray ionization),
calculated from equations (\ref{eq:uv_xray}) and the value of $J_{\rm crit}/J_{\rm crit,0}$
for $h\nu_{\rm min}=1$ keV and $T_\ast =3\times 10^4$ K shown in Fig.~\ref{fig:1}, 
in the two-dimensional 
parameter space ($J_{\rm crit,0}$, $f_{\rm X}/f_{\rm LW}$).
The shaded regions show the regions in this parameter space where the number density 
of DCBHs ($n_{\rm DCBH}$, calculated using equation \ref{eq:nDCBH}) falls below the value of 
$\sim 10^9~\msun$ SMBHs observed at $z\sim 6$.
The lightly shaded region shows this space for $n_{\rm DCBH}$ evaluated at $z=10$, 
corresponding to $J_{\rm crit} \ga 700$; the darkly shaded region shows where 
$n_{\rm DCBH}(z=20) < 1~{\rm cGpc}^{-3}$, 
corresponding to $J_{\rm crit} \ga 3\times 10^3$. 
Note that these values are conservative, because they ignore metal enrichment by 
\textit{in situ} star formation, as well as the decrease in the comoving number density 
that results from hierarchical merging of DCBH-forming halos. 
We therefore argue that the enhancement of $J_{\rm crit}$ by soft X-ray irradiation 
rules out this region in the physical parameter space. 
Most of the theoretical expectations discussed above 
(regarding $f_{\rm X}$, $f_{\rm LW}$ and $J_{\rm crit,0}$; 
\S\ref{subsec:actual_jcr}) point to $J_{\rm crit}$ being $> 10^3$.

Furthermore, we note that the requirement that the DCBH number density at $z\sim10$ should 
be greater than $\sim 1$ Gpc$^{-3}$ is itself also very conservative 
(in addition to the assumptions made regarding $P_{\rm gen}$). 
This is because this value reflects only the abundance of  $\sim 10^9~\msun$ SMBHs at $z\sim 6$. 
If one stipulates that the FUV DCBH scenario must also account for the observed $\sim 10^8~\msun$ 
SMBHs in the same redshift range, then the number density of DCBH must be at least 
$\sim 100$ Gpc$^{-3}$ \citep{2010AJ....139..906W}. 
All of these factors strongly put into question the viability of the FUV-aided DCBH scenario 
in explaining the $z\sim 6$ quasar observations.

\section{Discussion and Conclusions}
\label{sec:conc}

In this paper, we investigated the effect of X-ray irradiation on direct collapse black hole (DCBH) 
formation via far-ultraviolet (FUV) irradiation.
X-ray ionization promotes the H$_2$ formation because H$_2$ molecules are produced by 
the electron-catalyzed reactions (equation~\ref{eq:H2react2}).
Thus, X-ray irradiation increases the critical FUV flux $J_{\rm crit}$ required 
to suppress the H$_2$ formation and cooling.
Specifically, we focused on the effects of soft ($\sim 1~{\rm keV}$) X-rays 
emitted by the same galaxies close-proximity star-forming galaxies
that provide the large FUV intensities necessary for direct collapse.
Our main findings are as follows:
\begin{enumerate}
\item{\textbf{Galaxies supplying large $J_{\rm LW}$ should also provide a large soft X-ray intensity $J_{\rm X}$. }
If $z>10$ star-forming galaxies have the
same X-ray-to-FUV emission ratio as observed in their lower-redshift counterparts,
then an FUV intensity $J_{\rm LW,21}$ of several $10^3$ or higher would
be accompanied by a soft X-ray intensity $J_{\rm X,21}>0.01$ (equation \ref{eq:uv_xray}).
}
\item{\textbf{Such X-ray intensities increase $J_{\rm crit}$ by
promoting $\Hmol$ formation.}
Soft X-ray intensities ($J_{\rm X,21}>0.01$)
are sufficient to increase $J_{\rm crit}$
by a factor of a few or more, even under the conservative assumption that the X-ray-to-FUV
ratio does not evolve with redshift.
This finding agrees with that of \cite{Latif+14}, who considered a background
of hard $2-10$ keV X-rays.
Whereas $J_{\rm X,21}> 0.01$ is unlikely for a cosmic X-ray background,
it is plausibly the norm for putative DCBH sites.
}
\item{\textbf{This change in $J_{\rm crit}$ results in a drop of several orders of magnitude
in the DCBH abundance, compared to calculations that neglect the X-ray intensity.}
The point that $J_{\rm crit}$ may be too high (and DCBH sites
too rare) to account for the high-redshift quasar observations was previously raised
by \citetalias{Dijkstra+14}. Our results imply that accounting
for the soft X-rays produced locally---by the very same galaxies that supply
the photodissociating FUV photons necessary for direct collapse---places
an additional stress on the viability of the DCBH model to explain SMBH formation in general.
}
\end{enumerate}

We stress two points.
The first point is that this X-ray enhancement of $J_{\rm crit}$
is a general effect. 
Our calculations utilized several simplifying assumptions.
For one, we relied on one-zone estimates that do not include
three-dimensional effects (e.g. turbulence, anisotropy, and inhomogeneity)
and detailed hydrodynamics.
For another, our semi-analytic estimates of galaxy/SMBH number densities
at high redshifts (following \citetalias{Dijkstra+14})
suffer from uncertainties in the properties of the earliest stars and galaxies.
While improvements on these areas will surely lead to a more realistic quantitative
estimate of $J_{\rm crit}$ and $n_{\rm DCBH}$,
we expect our main findings enumerated above to be qualitatively robust.
This is precise the reason that in many parts of this paper we treat
the X-ray-\textit{uncorrected} critical FUV intensity $J_{\rm crit,0}$
(which could change with improved physical modeling,
as well as vary from one potential DCBH site to another)
as a loose parameter, and instead focus on
the X-ray enhancement factor $J_{\rm crit}/J_{\rm crit,0}$.

The second point of note is that this effect is highly likely to be astrophysically relevant.
We have adopted conservative parameter values in estimating the
X-ray intensity due to the FUV source galaxies.
As stated above: the X-ray intensity $J_{\rm X,21}\sim 10^{-2}$
required to affect $J_{\rm crit}$ (see Figs.~\ref{fig:1} and \ref{fig:2}),
while much higher than the cosmic X-ray background,
should be fairly typical for putative DCBH formation sites with $J_{\rm LW,21}>10^3$.
If direct collapse requires $J_{\rm LW,21}\sim 10^4$
(as suggested by the most recent simulations),
$J_{\rm X,21}$ would be $\sim 10^{-1}$.
The actual X-ray intensity could be still higher if star-forming galaxies
at $z>10$ produced more $\sim 1-2$ keV X-rays relative to FUV photons
(i.e. $f_{\rm X}/f_{\rm LW}>1$)---due to higher HMXB activity.

\subsection{Constraints from 21 cm observations}
We have shown that X-ray ionizations may play a powerful role in suppressing FUV-aided DCBH formation.
As shown in \S\ref{sec:prob} (Figs.~\ref{fig:n_DCBH} and \ref{fig:5}),
the predicted number density of 
DCBHs at $z\simeq 10-20$ for $f_{\rm X}\ga 1$ 
can be much less than that of the observed high-z QSOs with $M\sim 10^9~\msun$,
$\sim 1~{\rm comoving~Gpc}^{-3}$
(or $\sim 100~{\rm comoving~Gpc}^{-3}$ for $M\ga 10^8~\msun$ SMBHs).
For $f_{\rm X}\ga 0.1$, if $J_{\rm crit,0} \ga 2\times 10^3$,
FUV-DCBH formation is ruled out as it cannot make enough seed BHs
at $10<z<20$ to account for the observed $z\sim 6$ quasar population.
However, the possibility that $f_{\rm X}$ is much smaller than $0.1$ is not excluded 
by current observational data.

The 21-cm line transition of neutral hydrogen is one of the most promising observations
to probe the thermal history of the intergalactic medium before the cosmic reionizaion.
Future observations of 21-cm signals from the high-$z$ Universe could give 
a lower limit on $f_{\rm X}$ and thus help to further test the viability of the FUV-DCBH scenario.
The power spectrum of the brightness temperature of the 21-cm line
at the scale of $\simeq 0.1~{\rm Mpc}^{-1}$ has three peaks as a function of redshift
\citep[e.g.,][]{2007MNRAS.376.1680P,2013MNRAS.431..621M,2013JCAP...09..014C}.
The location and amplitude of the second peak 
is sensitive to the value of $f_{\rm X}$.
As the X-ray intensity is weaker than that from star-forming galaxies at the low-$z$ universe
($f_{X}\ll 0.1$), the peak position shifts toward lower-redshift and its amplitude becomes larger.
In this case, the 21-cm signal can be observed by 1st-generation interferometers; 
e.g. the Low Frequency Array (LOFAR, \citealt{2013A&A...556A...2V}) and 
Murchison Wide Field Array (MWA, \citealt{2013PASA...30....7T}).
On the other hand, for stronger X-ray intensities ($f_{\rm X}\gg 0.1$),
the spin temperature approaches the CMB temperature due to X-ray heating at higher redshift.
Then, the peak of 21-cm signal moves to higher redshift and becomes smaller.

Second generation interferometers such as the Square Kilometre Array (SKA, \citealt{2013ExA....36..235M}) 
will be required to observe the signal for larger $f_{\rm X}$.
However, near-future observations will be able to impose a lower limit of X-rays in the early Universe 
around $f_{\rm X}\sim 0.1$ \citep[e.g.,][]{2013JCAP...09..014C, 2014MNRAS.439.3262M}.
The same observations should also be able to constrain the efficacy of SMBH growth through 
rapid growth of Pop III remnants (Tanaka, O'Leary \& Perna, in prep.).

\subsection{Co-production of X-ray and FUV radiation}
In this paper, we assume that star-forming galaxies emit X-rays as well as FUV radiation.
This assumption seems reasonable when star formation occurs continuously during 
the cosmic time at the high-$z$ Universe ($\sim 300$ Myr for $z\sim 15$),
because the lifetime of massive stars is shorter than $\sim 10$ Myr.
The lifetime $t_{\rm life}$ of massive stars with mass of $15/25/40/200~\msun$ is 
$10/6.5/3.9/2.2$ Myr \citep{2002A&A...382...28S}.
Observations of star-forming galaxies at lower redshift also support this assumption 
\citep[e.g.,][and references therein]{2012MNRAS.419.2095M}.
The co-production of X-rays with FUV photons is also seemingly unavoidable when a galaxy 
or an atomic-cooling halo undergoes an intense and short burst of star formation. 
Because some fraction of the newly formed stars will die and form HMXBs, 
there may be a very narrow window (at most $t_{\rm life} \sim$ a few Myr) inside 
which a DCBH forming halo is irradiated by FUV radiation but not X-rays.

Let us also consider the gas properties inside a DCBH-forming halo. 
The gas density at the central core, before radiative cooling operates, 
is $\la 10~\cc~((1+z)/11)^{3}$ \citep{Visbal+14}.
After the virial temperature reaches $\ga 8000$ K, H atomic cooling causes 
the gas to undergo gravitational collapse. 
Since the density increases on the free-fall timescale, it will take 
$t_{\rm coll}\ga 20$ Myr $((1+z)/11)^{-3/2}$ for the gas density to exceed $\sim 10^3~\cc$-- 
the value where H$_2$ molecules can be collisionally dissociated instead of by FUV irradiation. 
Then it follows that X-ray irradiation must accompany any strong FUV intensity 
as long as $t_{\rm life}<t_{\rm coll}$.

Recently, \cite{2014MNRAS.445.1056V} proposed a new DCBH formation scenario, 
which considers synchronised pairs of pristine atomic 
cooling halos having a small separation $\la 0.5$ kpc.
If one of the halos reaches the atomic cooling threshold (i.e., $T_{\rm vir}\ga 8000$ K) 
just after star formation occurs in another halo, the first halo can be irradiated with 
the critical FUV intensity due to a small separation
(SFR $\sim 0.05~\msunyr$ is required to realize $J_{\rm crit,0}\sim 2\times 10^3$).
To be viable, this scenario must keep the gas free of ionizing X-rays and metal-enriching 
winds for a collapse timescale $\sim 20$ Myr $((1+z)/11)^{-3/2}$. 
As argued above, the FUV-producing massive stars can become X-ray sources on 
a significantly shorter timescale $t_{\rm life}$. 
Thus, we argue that X-ray irradiation can also suppress this``synchronised pair" scenario. 
(Also note that the timescale on which winds can reach and pollute the gas is also 
comparable to $t_{\rm coll}$; \citetalias{Dijkstra+14}.) 
However, these arguments (lifetime of massive stars, photo-evaporation, 
and metal pollution) depend on a number of uncertain parameters 
(e.g. initial mass function, star formation efficiency, clumping factor of the intergalactic medium, 
and wind velocity).
We also note that the X-ray luminosity estimated from equation (\ref{eq:LX_SFR}) is 
$\sim 3.4\times 10^{38}~({\rm SFR/0.05~\msunyr})$ erg s$^{-1}$, which is a few times 
the Eddington luminosity of stellar mass BH.
This fact implies that only a handful of luminous X-ray binaries are required to achieve 
$J_{\rm crit,0}$ in this particular case, and that X-ray luminosities in such low-mass halos 
may have a dispersion larger than the $\sim 0.4$ dex value found in more massive galaxies.
A finer knowledge of these details will be required to better understand 
the impact of various photon sources inside such close, synchronised halo pairs.

\subsection{Other effects to enhance $J_{\rm crit}$ and suppress the DCBH formation}
We here discuss the suppression of DCBH formation by other ionization effects.
In the early universe, the promising ionizing radiations other than X-rays are cosmic rays (CRs)
and EUV photons ($\geq 13.6$ eV) from star-forming galaxies.
These sources of radiation
increase the ionization fraction of the gas in the atomic cooling halos where 
SMSs would be born.
Thus, they should also increase the requisite FUV intensity for
DCBH formation, in much the same way as the effect of X-ray ionizations
discussed in this work.

The enhancement of $J_{\rm crit}$ by CR ionization operates 
when the ionization rate is larger than $\sim 10^{-18}$ s$^{-1}$ at
H column densities of $\sim 10^{22}$ cm$^{-2}$, which corresponds to
$n\sim 10^3~\cc$ and $T\sim 8000$ K \citep{IO11}.
Assuming the CR energy distribution comprises a power-law spectrum of 
${\rm d}n_{\rm CR}/{\rm d}E\propto E^{-2}$ with $10^6\leq E \leq 10^{15}$ eV,
the ionization rate required to increase $J_{\rm crit}$ can be estimated as 
$\zeta_{\rm CR}\ga 10^{-17}$ s$^{-1}$,
which is smaller than that observed in Milky Way, $10^{-17}\la \zeta_{\rm CR}\la 10^{-15}$ s$^{-1}$
\citep{1961PASJ...13..184H, 1968ApJ...152..971S, 
1998ApJ...506..329W, 2003Natur.422..500M, 2007ApJ...671.1736I}.
According to theoretical estimate \citep[e.g.][]{2007MNRAS.382..229S, IO11, 2014MNRAS.442.2667N}, 
the CR ionization rate is 
\begin{equation}
\zeta_{\rm CR}\sim 2\times 10^{-20}{\rm s}^{-1}\left(\frac{d}{10~{\rm kpc}}\right)^{-2}\left(\frac{{\rm SFR}}{\msunyr}\right),
\label{eq:CR}
\end{equation}
where we assume that 10 per cent
of the supernovae explosion energy ($E_{\rm SN}=10^{51}$ erg) 
converts to the CR acceleration and the Salpeter initial mass function
as we assumed for the FUV intensity.
Combining equation (\ref{eq:CR}) and $J_{\rm LW,21}=f_{\rm LW}{\tilde J_{\rm LW,21}}$, then we can find 
\begin{equation}
\zeta_{\rm CR}\sim 2\times 10^{-19}{\rm s}^{-1}~f_{\rm LW}^{-1}\left(\frac{J_{\rm LW,21}}{10^3}\right),
\end{equation}
which is smaller than the ionization rate above which $J_{\rm crit}$ increases.
However, the CR intensity in the early universe has uncertainties associated with magnetic fields,
e.g., the confinement of CRs in star-forming galaxies and CR propagation in the intergalactic medium
\citep[e.g.][and references therein]{2007ARNPS..57..285S}.
Thus, more sophisticated models are required to better evaluate the impact of CR ionizations
on DCBH formation.

EUV photons with $\geq 13.6$ eV can easily absorbed by the intergalactic medium 
because their optical depth in neutral hydrogen is as large as
\begin{equation}
\tau _{\rm H}(\nu)\sim 10^4\left(\frac{n}{1.0~\cc}\right)^{1/2}\left(\frac{h\nu}{13.6~{\rm eV}}\right)^{-3}.
\end{equation}
Thus, EVU photons cannot penetrate into the dense and hot region 
($n\sim 10^3~\cc$ and $T\simeq 8000$ K) in atomic cooling halos.
As a result, the critical FUV intensity does not change because of self-shielding to EUV photons.
However, EUV photons would suppress DCBH formation during and after cosmic reionization,
only when the gas around atomic-cooling halos becomes ionised completely 
\citep{2014MNRAS.440.1263Y, 2014MNRAS.445..686J}.

\subsection{Alternative models forming DCBHs}
We discuss alternative scenarios of forming DCBHs which do not require strong FUV radiation.
The relevant H$_2$ dissociating process instead of FUV photodissociation is collisional dissociation 
(H$_2$ + H $\rightarrow$ 3H).
In atomic cooling halos ($T_{\rm vir}\ga 10^4$ K), this process works efficiently in case that the gas density 
and temperature are $n\ga 10^4~\cc$ and $T\ga 6000$ K, respectively.
Once the primordial gas enter such a dense and hot region (so-called ``zone of no return"), 
the H$_2$ formation/cooling is quenched by the collisional dissociation even {\it without FUV radiation} 
enough for the gas to collapses keeping a high temperature ($\sim 8000$ K) by H-atomic cooling \citep{IO12}.

One promising process for forming such a dense and hot gas is strong shock 
by collisions of cold accretion flows due to assembly of the first galaxies.
Since the radiative cooling of the gas is efficient in the first galaxies, 
the gas can penetrate deep to the centre ($\sim 0.1~R_{\rm vir}$) 
as dense filamentary inflows.
If the cold flows jump into the zone of no return by shock heating, 
a supermassive star can form from the parent cloud in the post-shock region.
However, supersonic filamentary flows are unlikely to be dense before 
experiencing shocks for weak-cooling case 
($T_{\rm vir} \la 8000$ K; \citealt{2014MNRAS.439.3798F}) and no-cooling case \citep{Visbal+14}.
These two examples suggest that massive halos ($T_{\rm vir}\ga 10^4$ K) could be necessary for the gas 
to arrive in the zone of no return.
To better understand the actual probability of a dense shocked gas cloud forming in this way, 
a large, statistical sample of numerical simulations of atomic-cooling halos is required.

A galaxy merger is another mechanism that can induce strong inflows and form an environment 
similar to the one made by the cold accretion shocks.
\cite{2010Natur.466.1082M} performed a numerical simulation of the merger between 
massive ($\sim 10^{12}~\msun$) and metal-enriched ($\sim \zsun$) protogalaxies at $z\sim 6$,
assuming a simple polytropic equation of state (i.e., $p\propto \rho ^\gamma$).
After the merging, the gravitationally unstable disc is formed, where 
the non-axisymmetric structures (spiral arms and bars) transport the gas angular momentum efficiently
In that case, strong inflows rapidly accumulate a mount of gas with $10^8~\msun$ within the central pc scale.
The average density of the nuclear region reaches $\sim 10^9~\cc$, 
at which point H$_2$ molecules can remain collisionally dissociated.
However, \cite{2013MNRAS.434.2600F} have noted that the gas actually fragments into clumps 
with $\la \msun$ if one considers more realistic radiative cooling prescriptions,
instead of a simple equation of state.
\cite{2010MNRAS.409.1022V} discussed the possibility that gas can accumulate in the galaxy center due to rapid angular momentum transfer via bar instabilities, as long as the mass inflow rate is higher than the star formation rate.
Further research is required to determine whether the galaxy merger could produce 
massive clouds forming DCBHs, 
when fully accounting for the cooling and chemical reactions of primordial gas.

A third avenue for forming DCBHs without FUV radiation was recently proposed by \cite{TL14}. 
The relative bulk streaming motion between baryons and dark matter left over from 
cosmic recombination \citep{2010PhRvD..82h3520T} has been shown to delay gas infall and 
Pop III star formation in $z \ga 20$ halos with $T_{\rm vir}\sim 1000-2000$ K 
\citep{2011ApJ...730L...1S, 2011ApJ...736..147G, 2012MNRAS.424.1335F, 2013ApJ...763...27N}. 
\citeauthor{TL14} noted that in rare combinations of particularly massive halos and 
exceptionally large streaming velocities, the delay in gas infall may persist until the 
halo reaches $T_{\rm vir} \sim 8000$ K. 
Gas falling into such halos would naturally shock to $8000$ K before ever forming stars.
The gas will undergo direct collapse if it can reach sufficiently large densities to keep 
H$_2$ collisionally dissociated (however, note the caveats and uncertainties discussed above). 
\cite{TL14} predicted that this this mechanism a characteristic redshift $z\sim 30$, 
where the product of the atomic-cooling halo number density 
and the probability of having a sufficiently large streaming velocity 
(i.e. significant delay in gas infall) is maximized.

\subsection{The Effect of Metallicity}

We briefly consider the possibility that the effect discussed in this paper
could be alleviated by the absorption of X-rays by heavy elements
(present in the interstellar or circumgalactic medium of the FUV/X-ray source galaxy).
Indeed, the total cross section of metals in the interstellar medium is larger than 
those of H and He by $1-2$ orders of magnitude at $1-10$ keV, 
at solar abundances \citep[e.g.][]{1983ApJ...270..119M}.
However, the metallicities in the environments of interest here are likely to be much lower than solar.
For $Z\la 10^{-3}~\zsun$, the contribution of metals to the optical depth is less
than 10 per cent, and absorption by metals is unlikely to shield the putative DCBH
formation sites from X-rays.

\section*{Acknowledgements}
We thank Mark Dijkstra, Zolt\'an Haiman, Eli Visbal, Kazuyuki Omukai 
and Kazuyuki Sugimura for fruitful discussions, 
as well as Jarrett L. Johnson, Muhammad Latif, Dominik Schleicher, and 
Marta Volonteri for comments on the manuscript.
We also thank Akio Inoue for providing the data of the spectral model of Pop II galaxies. 
This work is partially supported by Grants-in-Aid from the Ministry of Education, 
Culture, and Science of Japan (to KI).

{\small \bibliography{ref.bib}}

\appendix
\section{Increase of the Electron Fraction by X-rays}

\begin{figure*}
\begin{tabular}{cc}
\begin{minipage}{0.5\hsize}
\begin{center}
\includegraphics[height=60mm,width=80mm]{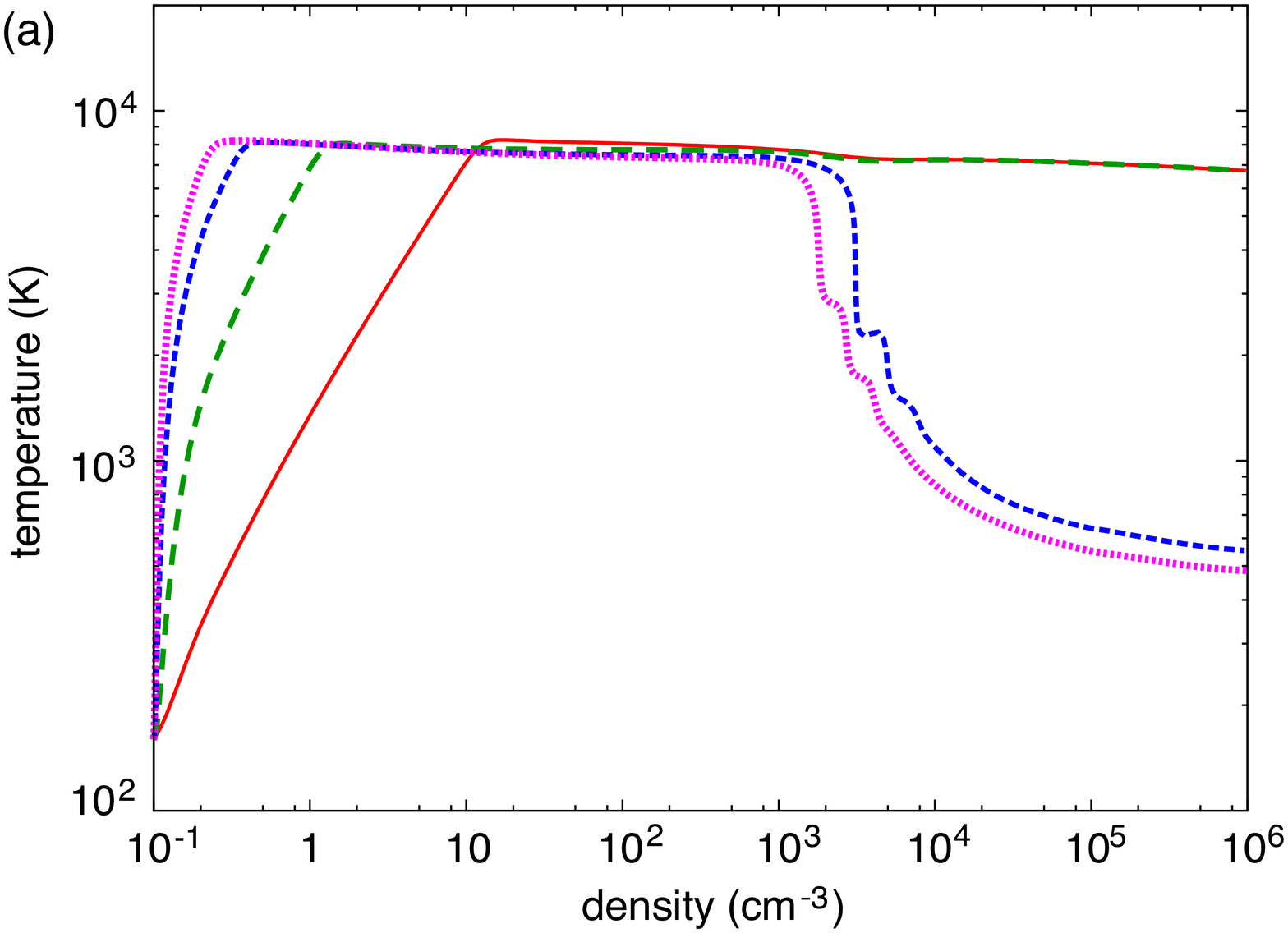}
\end{center}
\end{minipage}
\hspace{-5mm}
\begin{minipage}{0.5\hsize}
\begin{center}
\includegraphics[height=60mm,width=80mm]{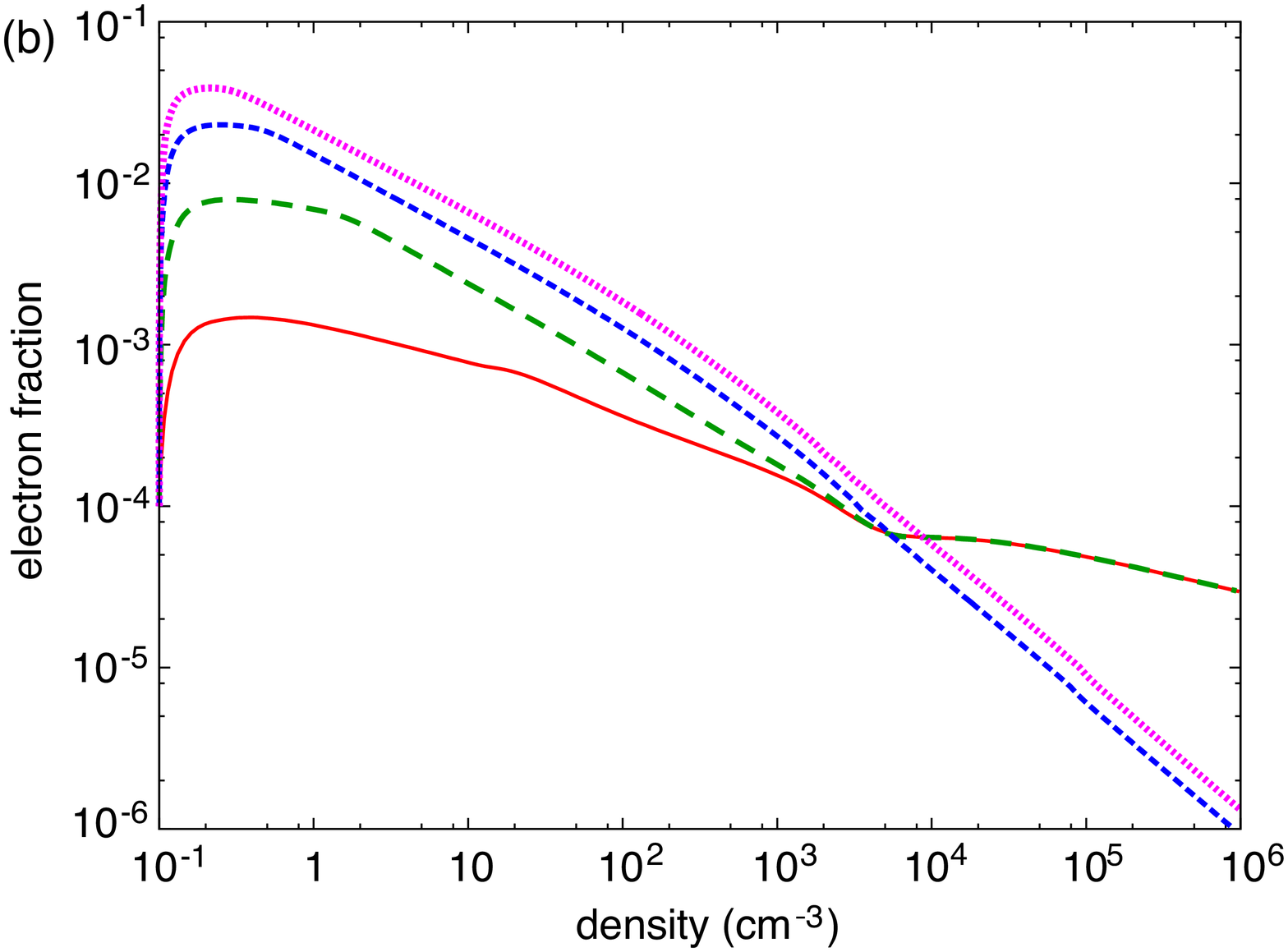}
\end{center}
\end{minipage}
\\
\\
\begin{minipage}{0.5\hsize}
\begin{center}
\includegraphics[height=60mm,width=80mm]{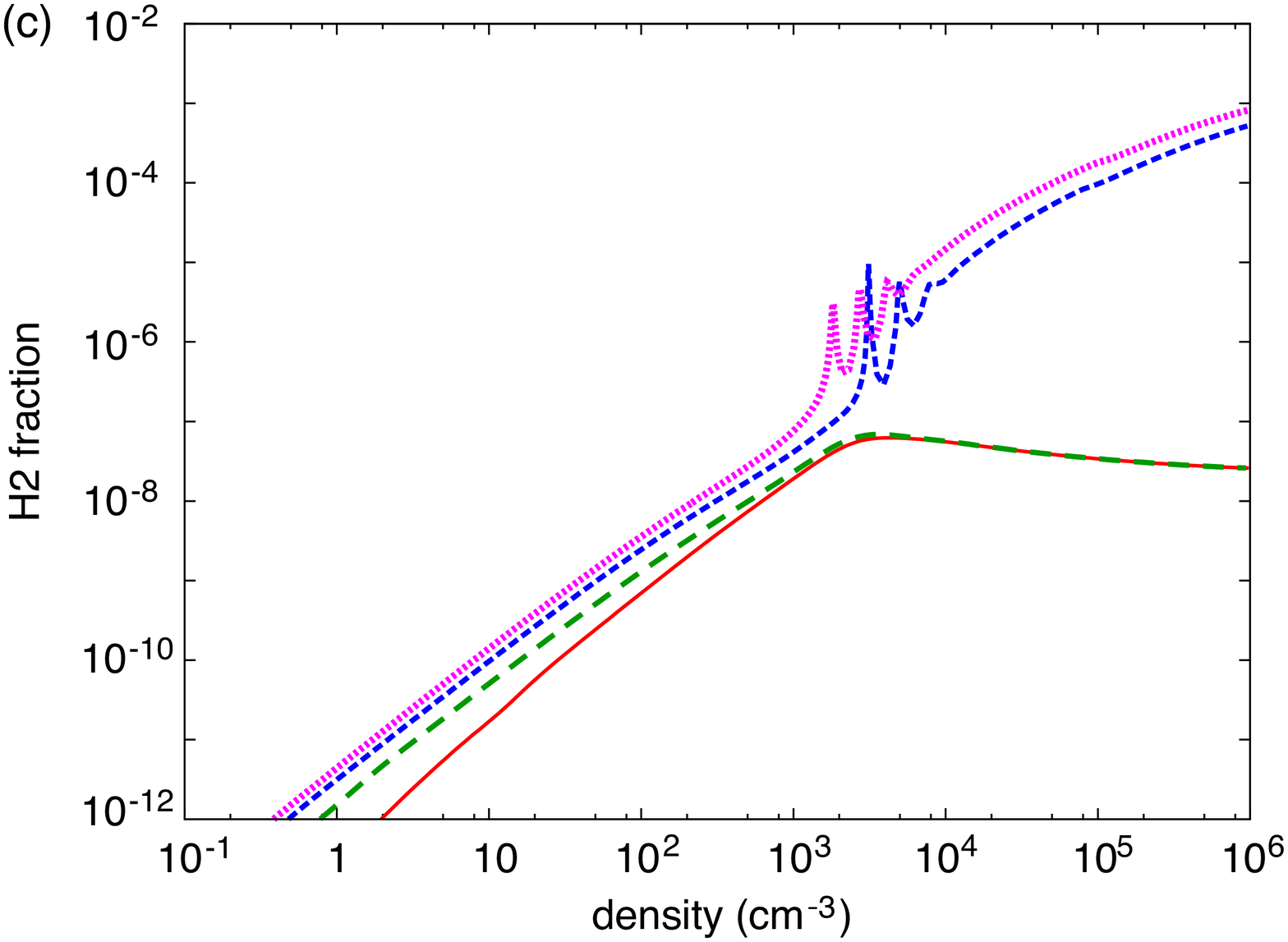}
\end{center}
\end{minipage}
\hspace{-5mm}
\begin{minipage}{0.5\hsize}
\begin{center}
\includegraphics[height=60mm,width=80mm]{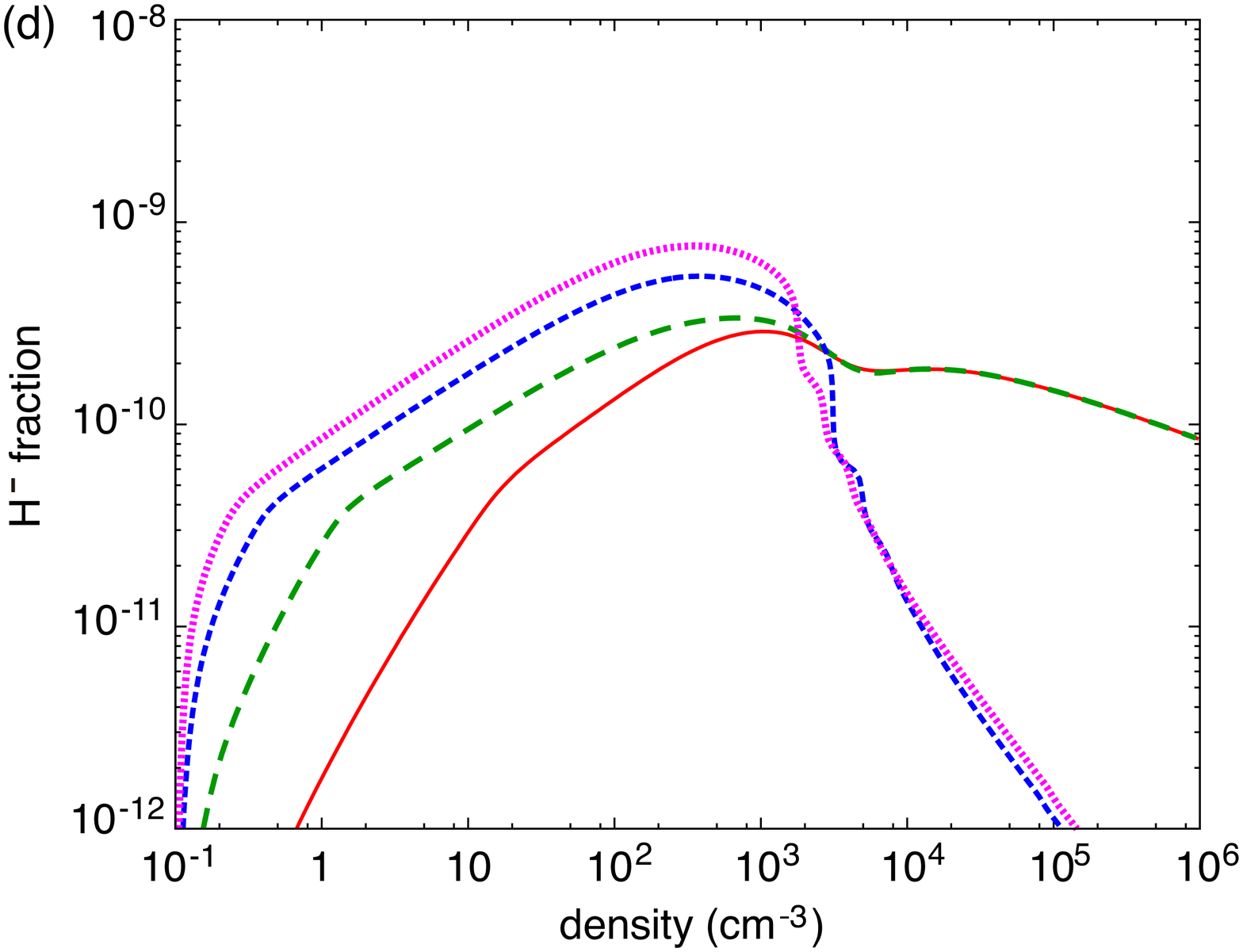}
\end{center}
\end{minipage}
\end{tabular}
\caption{Evolution of the (a) temperature, (b) electron fraction,
(c) H$_2$ fraction, and (d) H$^-$ fraction in a collapsing cloud
irradiated by FUV radiation with $J_{\rm LW,21}=10^4$ for $T_\ast =3\times 10^4$ K
and X-rays with $J_{\rm X,21}=10^{-3}$ (red solid), $10^{-2}$ (green long-dashed), 
$4.4\times 10^{-2}$ (blue short-dashed) and $10^{-1}$ (purple dotted).
A colour version of this figure is available in the online version.}
\label{fig:evolve}
\end{figure*}

In this Appendix, we discuss the effect of X-ray irradiation on thermal evolution of 
a collapsing gas cloud via the enhancement of the electron fraction $x_{\rm e}$.
This is a key underlying point of this paper.

Each panel of Fig.~\ref{fig:evolve} shows the temperature (a), electron fraction (b),
H$_2$ fraction (c), and H$^-$ fraction (d) as a function of the number density, respectively.
To see the effect of X-ray ionization, we first fix the FUV intensity to $J_{\rm LW,21}=10^4$ 
with a brightness temperature of $3\times 10^4$ K.
Each line represents the case of $J_{\rm X,21}=10^{-3}$ (solid), $10^{-2}$ (long dashed),
$4.4\times 10^{-2}$ (short dashed), and $10^{-1}$ (dotted).
We note that the case $J_{\rm X,21}=4.4\times 10^{-2}$
corresponds to our fiducial case for  $J_{\rm LW,21}=10^4$
(i.e. $f_{\rm X}/f_{\rm LW}=1$ in  equation~\ref{eq:uv_xray}).
The dependence of $J_{\rm crit}$ on $J_{\rm X}$ can be found in
Fig.~\ref{fig:2}, with the middle (green) solid line representing
the case $T_\ast =3\times 10^4$ K discussed in this Appendix.

From Fig.~\ref{fig:evolve} (a), we see how the X-ray intensity affects thermal evolution. 
For the weakest X-ray case ($J_{\rm X,21}=10^{-3}$; red solid curve),
the temperature increases almost adiabatically with the gas density ($T\propto n^{2/3}$).
After heating up to $10^4$ K, the gas collapses isothermally ($\simeq 8000$ K) via atomic-hydrogen cooling.
With stronger X-ray intensities, at low densities ($\la 10~\cc$)
the temperature increases more rapidly due to X-ray photoheating.
At higher densities ($\ga 10^3~\cc$), however, the gas cools down to $\sim 10^3$ K
for $J_{\rm X,21}=4.4\times 10^{-2}$ (blue short-dashed curve) and $J_{\rm X,21}=0.1$ (purple dotted).
The sudden drop of the temperature is caused by H$_2$ cooling, which is promoted by X-ray ionization 
through the electron-catalyzed reactions (eqs. \ref{eq:H2react1} and \ref{eq:H2react2}).

Fig.~\ref{fig:evolve} (b) shows that the electron fraction $x_{\rm e}$ decreases as the cloud collapses,
with radiative recombination balancing X-ray ionization.
For the cases with weaker X-rays (red solid and green long-dashed), 
at large densities $n\ga 10^4~\cc$
the gas keeps a high temperature 
of $\simeq 8000$ K, with  $x_{\rm e}$ eventually converging to a value $\simeq 4\times 10^{-5}$.
For stronger X-ray intensities(blue short-dashed  and purple dotted), the electron fraction increases by X-ray ionization
past $n\sim 10^3~\cc$.
This is roughly the density value at which the bifurcation of thermal evolution is determined---i.e., 
whether the gas remains $\Hmol$-free and nearly isothermal at $\sim 8000$ K, or forms $\Hmol$ and cools.
Due to the enhancement of $x_{\rm e}$, more H$_2$ molecules (as well as H$^-$ ions) form 
before the bifurcation point (Fig.~\ref{fig:evolve} c and d).

\begin{figure}
\begin{center}
\includegraphics[height=60mm,width=80mm]{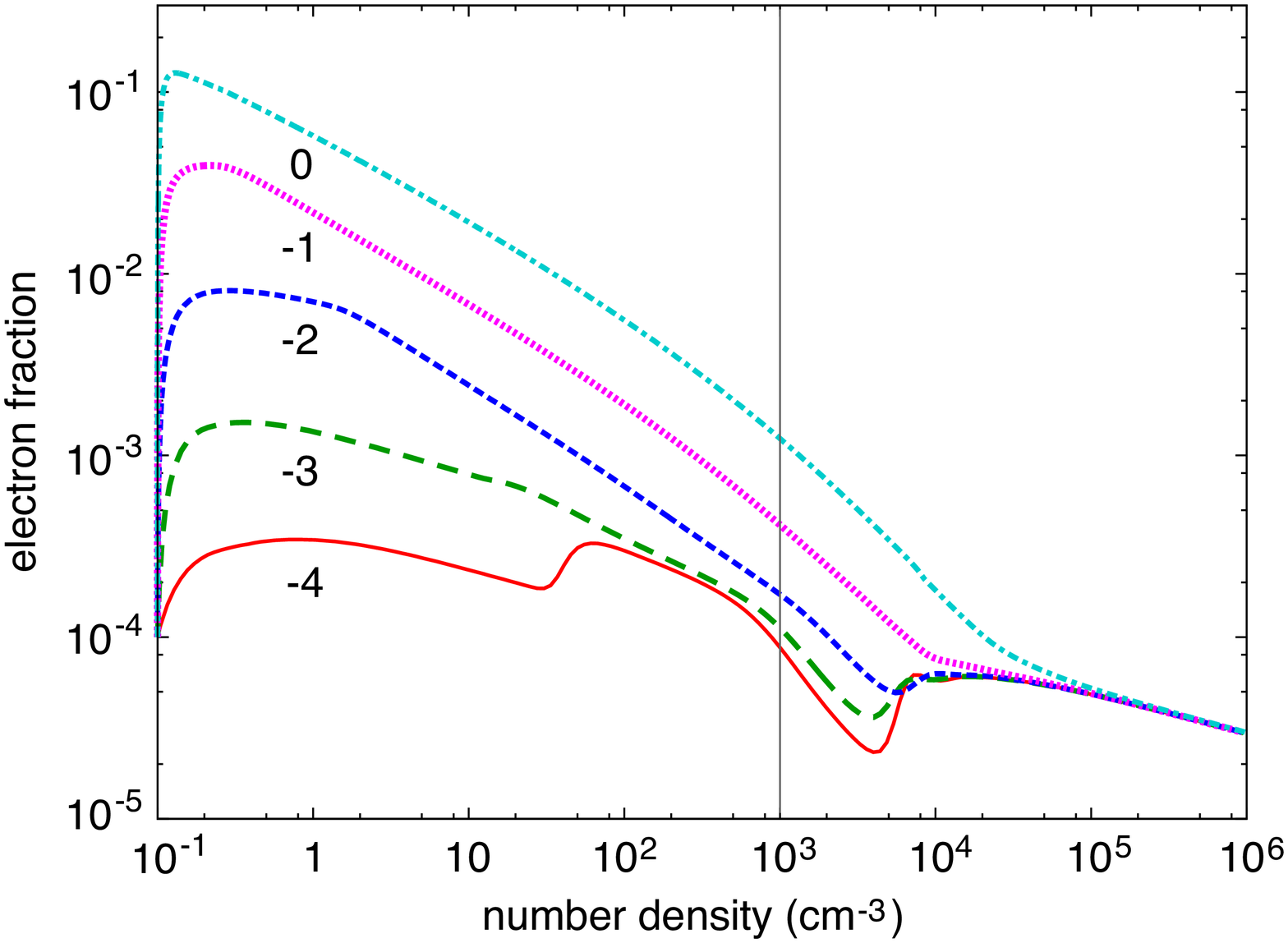}
\end{center}
\caption{Evolution of the electron fraction in a collapsing cloud
for ${-4}\leq \log J_{\rm X,21}\leq 0$, the values of which are denoted by numbers in the figure.
For each case, the FUV intensity $J_{\rm LW,21}$ is set slightly higher 
than the critical value $J_{\rm crit}$ for $T_\ast =3\times 10^4$ K (see Fig.~\ref{fig:2}).
A colour version of this figure is available in the online version.}
\label{fig:x_e}
\end{figure}

Next, we clarify the dependence of that the critical FUV on the X-ray intensity 
(i.e. $J_{\rm crit}\propto J_{\rm X,21}^b$; $b\simeq 0.5$).
In Fig.~\ref{fig:x_e}, we show the evolution of the electron fraction again,
for X-ray intensities $J_{\rm X,21}=10^{-4}$, $10^{-3}$, $10^{-2}$, $10^{-1}$ and $1$ (from the bottom to the top).
For each value of $J_{\rm X,21}$ represented in the figure,
we have set the FUV intensity $J_{\rm LW,21}$ slightly higher than the critical value 
$J_{\rm crit}$ (accounting for the enhancement in this value due to X-ray ionizations)
for $T_\ast =3\times 10^4$ K (see Fig.~\ref{fig:2}):
$J_{\rm LW, 21} \approx 2.5\times 10^3$, $3.5\times 10^3$, $7\times 10^3$, $2.5\times 10^4$,
and $8\times 10^4$ (from the bottom to the top).
For all cases shown in this figure,
the temperature evolution is nearly isothermal at $T\sim 8000$ K
once the gas density reaches $n\ga 30~\cc$.

For all cases with $J_{\rm X,21}\la 10^{-2}$ in Fig.~\ref{fig:x_e}, 
the electron fraction is roughly the same at $n\simeq 10^3~\cc$, 
where the bifurcation of thermal evolution occurs (vertical solid line).
Thus, the corresponding values of $J_{\rm crit}$ does not vary very much below this X-ray intensity value.
However, for $J_{\rm X,21}\ga 10^{-2}$, the electron fraction remains high even at $n\simeq 10^3~\cc$.
Therefore, if $J_{\rm X,21}\ga 10^{-2}$,
stronger FUV intensities are required to keep the gas $\Hmol$-free---in other words, $J_{\rm crit}$ increases.

\begin{figure}
\begin{center}
\includegraphics[height=60mm,width=80mm]{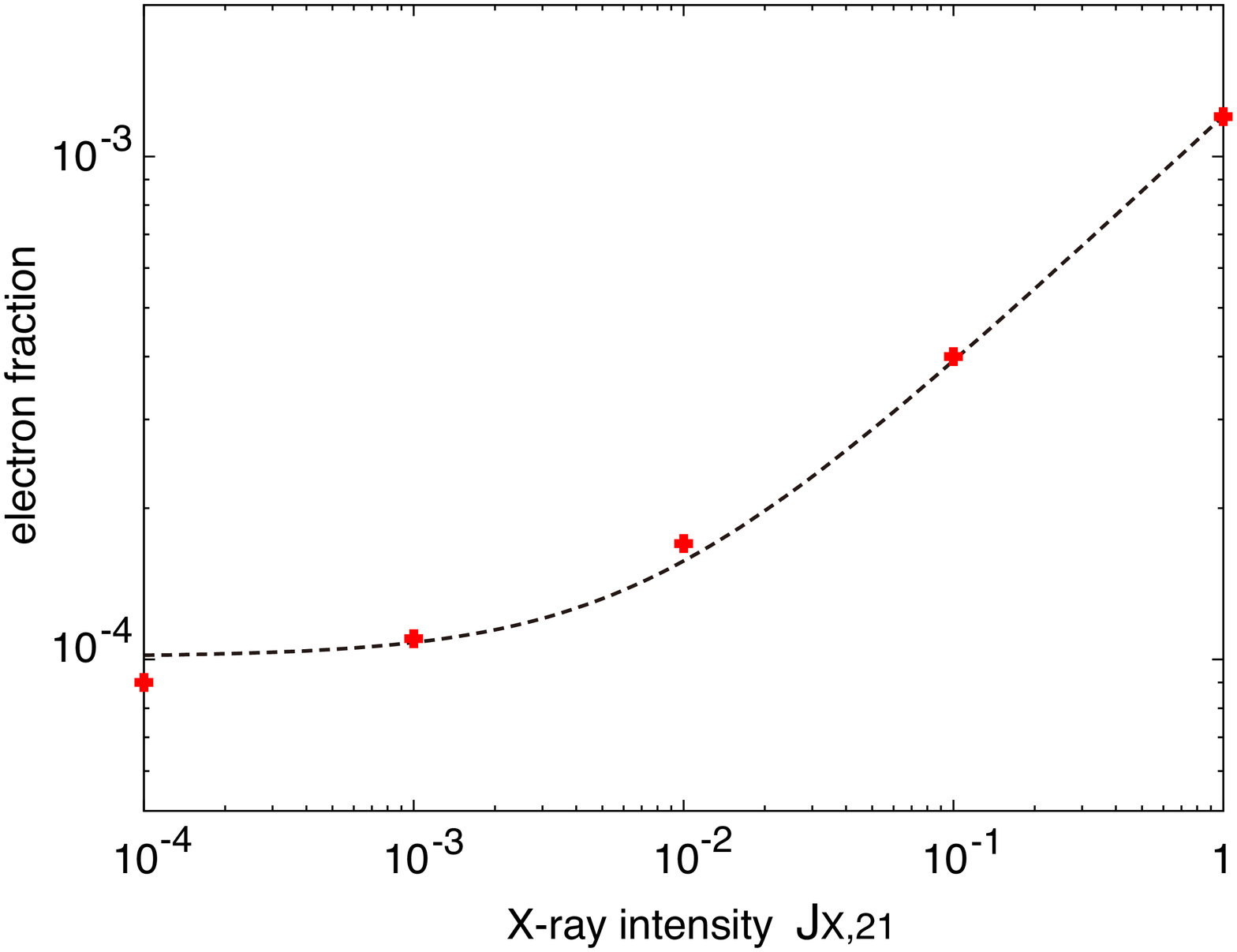}
\end{center}
\caption{Relation between the X-ray intensity and the electron fraction 
at $n=10^3~\cc$ (red cross symbols), i.e. the values at the vertical solid line in Fig.~\ref{fig:x_e}.
The dashed line is $10^{-4}(1+J_{\rm X,21}/6.7\times 10^{-3})^{0.5}$.
A colour version of this figure is available in the online version.}
\label{fig:Jx_xe}
\end{figure}

Finally, Fig.~\ref{fig:Jx_xe} shows the relation between the X-ray intensity and the electron fraction 
at $n=10^3~\cc$, i.e. the values on the vertical solid line of Fig.~\ref{fig:x_e}.
These numerical results (cross symbols) are explained well by a function of 
$10^{-4}(1+J_{\rm X,21}/6.7\times 10^{-3})^{0.5}$ (dashed line).
For large X-ray intensities $J_{\rm X,21} \ga 10^{-2}$,
$x_e$ is roughly proportional to $J_{\rm X}^{0.5}$;
at lower intensities, X-rays do not have an appreciable effect on $x_e$.
Previous works have shown that $J_{\rm crit}$ increases linearly with $x_e$
for hard FUV spectra \citep[e.g.,][]{O01,IO11}.
The above two relationships lead to the dependence of $J_{\rm crit}$ on $J_{\rm X}$
shown in this paper, and motivate the fitting formulae used in \S\ref{sec:2.3}.

\end{document}